\documentclass[pra,twocolumn,showpacs,floatfix]{revtex4}
\usepackage{amsmath,amsfonts,amssymb,amsthm,graphicx,epsfig,color,times,psfrag,subfigure}

\def\one{\ensuremath{\hbox{$\mathrm I$\kern-.6em$\mathrm 1$}}}

\newcommand{\tr}{\operatorname{tr}}
\def\tr{ \mbox{tr}}

\newcommand{\ket}[1]{\mbox{$| #1 \rangle$}}
\newcommand{\braket}[2]{\mbox{$\langle #1  | #2 \rangle$}}
\newcommand{\proj}[1]{\mbox{$|#1\rangle \!\langle #1 |$}}

\def\figurewidth{.19\textwidth}
\def\onefigurewidth{.38\textwidth}

\begin{document}

\title{Geometric Phases and Critical Phenomena in a Chain of Interacting Spins}

\author{Moritz E.\ Reuter}\email{moritz.reuter@imperial.ac.uk}
\author{Michael J.\ Hartmann}
\author{Martin B.\ Plenio}

\affiliation{QOLS, Blackett Laboratory, Imperial College London,
Prince Consort Road, London SW7 2BW, UK}

\affiliation{Institute for Mathematical Sciences, Imperial College
London, 53 Exhibition Road, London SW7 2BW, UK}

\date\today

\begin{abstract}
The geometric phase can act as a signature for critical regions of
interacting spin chains in the limit where the corresponding
circuit in parameter space is shrunk to a point and the number of
spins is extended to infinity; for finite circuit radii or finite
spin chain lengths, the geometric phase is always trivial (a
multiple of $2\pi$). In this work, by contrast, two related
signatures of criticality are proposed which obey finite-size
scaling and which circumvent the need for assuming any unphysical
limits. They are based on the notion of the \emph{Bargmann
invariant} whose phase may be regarded as a discretized version of
Berry's phase. As circuits are considered which are composed of a
discrete, finite set of vertices in parameter space, they are able
to pass directly \emph{through} a critical point, rather than
having to circumnavigate it. The proposed mechanism is shown to
provide a diagnostic tool for criticality in the case of a given
non-solvable one-dimensional spin chain with nearest-neighbour
interactions in the presence of an external magnetic field.
\end{abstract}

\maketitle

\section{Introduction}
\label{section:Introduction}
\subsection{Quantum Phase Transitions}
\label{subsection:QPTs}
Unlike ordinary phase transitions, quantum phase transitions
(QPTs)~\cite{Vojta} take place at a temperature of absolute zero
and are consequently driven entirely by quantum fluctuations,
rather than thermal ones. Moreover, the long-range, algebraically
decaying correlations which are characteristic of critical
many-body ground states are due entirely to entanglement. As such,
one would expect entanglement measures (see
Ref.~\cite{EntMeasures} for an overview) to provide further
insight into the fundamental physical underpinnings of QPTs,
possibly complementing the conventional condensed-matter approach,
which relies mostly on two-point correlation functions.

Indeed, Osterloh and co-workers~\cite{OsterlohEtAl} demonstrated
singular and scaling behaviour of a bipartite entanglement measure
in the vicinity of the critical point in the one-dimensional XY
spin model. On the other hand, this behaviour could be understood
entirely from that of the relevant two-point correlation functions
as these are sufficient to determine the two-particle reduced
density matrices. Furthermore, the entanglement between just two
sites is not particularly well suited to capturing the large-scale
behaviour of correlations, which becomes all the more relevant in
the critical regime. This was recognized in
Ref.~\cite{BlockEntIC}, where the scaling behaviour of the
entanglement between a contiguous \emph{block} of $L$ sites with
the rest of the lattice was first considered in lattice field
theories. It was found that the entanglement entropy of
\emph{noncritical} chains with short-ranged interactions reaches a
saturation level, while the entanglement entropy diverges
logarithmically with $L$ in the field limit, i.e. in the case of
\emph{critical} chains. This approach was subsequently pursued
numerically~\cite{VidalLatorre} in the particular context of the
one-dimensional XY model and, in more generality, in an analytical
treatment based on random matrix theory~\cite{Keating}, applicable
to any spin chain Hamiltonian that can be cast into a quadratic
form of fermionic operators.

At the same time, progress has been hampered by the realization
that \emph{bi-partite} entanglement, whether of blocks of spins or
otherwise, at best only gives us an incomplete, local picture of
the entanglement exhibited by generic many-body ground states. It
is something of a truism, of course, to note that a truly global,
multi-partite approach is indispensable if one wishes to capture
the entanglement that pervades critical many-body ground states
simultaneously at all length scales (for recent work on
multi-partite entanglement in quantum spin chains see for example
Refs.~\cite{Bruss,Costantini}).

Unfortunately, the theory of multi-partite entanglement is still
very much in its infancy. It is true that many of the entanglement
measures used for bi-partite states carry straightforward
generalizations to the multi-partite setting. This is particularly
true for distance-based entanglement measures, such as the
relative entropy of entanglement~\cite{RelEnt} or the closely
related geometric entanglement~\cite{Geometric}: the entanglement
of a state is then simply quantified in terms of the minimum
distance of that state from the set of all multi-partite separable
states, rather than the set of all bi-partite separable states.
However, it is one thing to generalize the axiomatic definitions
of entanglement measures to include the multi-partite case, but
quite another to still be able to compute these in practice.
Moreover, the theory of multi-partite entanglement is still
plagued by a host of different candidates for suitable
entanglement measures, even for the case of pure states (see
Ref.~\cite{EntMeasures} for more details).

In view of these difficulties, attention has shifted to include
other, potentially related, means of characterizing
QPTs~\cite{Zanardi}. One such approach centres around the notion
of \emph{geometric
phase}~\cite{Berry,Resta,Carollo1,Carollo2,Hamma}, and it is this
approach which we wish to pursue in the remainder of this work.

\subsection{Geometric Phase}
\label{subsection:GeometricPhase}
In his seminal paper~\cite{Berry}, Berry investigated the phase
picked up by an eigenstate of a parameter-dependent Hamiltonian
when transported adiabatically around a closed trajectory in
parameter space. It turns out that in addition to the well-known
dynamical phase there is also a geometric component to the phase,
which is observable, at least in principle. While the dynamical
phase provides a measure of the duration of the Hamiltonian's
evolution and is independent of the geometry of the trajectory
followed, conversely, the geometric phase is independent of the
rate at which the state is transported around the loop (as long as
this is slow enough for the adiabatic theorem to rule out
transitions to neighbouring, orthogonal states) and depends solely
upon the geometry of the trajectory. The geometric phase had
hitherto been widely overlooked as just another unphysical phase
factor, and certainly had not been granted the level of
recognition it enjoys nowadays.

Before going on to explain how Berry's phase may be used to probe
for QPTs, it is useful here to first restate a simple example
given in Berry's original paper, which serves to illustrate the
concept of geometric phase. To that end, consider a single
spin-$1/2$ particle which is coupled to an external magnetic
field, and suppose the spin is initially aligned with the magnetic
field, i.e. the particle is in an eigenstate of the system's
Hamiltonian. Then the spin will remain aligned with the magnetic
field vector (the particle remains in the instantaneous
eigenstate) when the magnetic field vector is made to rotate
adiabatically. Upon completion of some closed trajectory, the
final state of the particle differs from its initial one by an
overall phase factor, which is part dynamical, part geometric in
origin. Now, it turns out that the geometric phase is in fact
equal in size to precisely half the solid angle subtended at the
origin by the magnetic field vector's trajectory in parameter
space. This result not only serves to highlight in a particularly
acute way the geometric character of Berry's phase, but it also
gives an indication of the origin of Berry's phase: the spin's
state is degenerate when the magnetic field is switched off, and
the geometric phase can thus be interpreted as providing a measure
of the view of the circuit as seen from that point of degeneracy
at the origin of the parameter space.

In fact, in any general setting, the emergence of geometric phases
can always be traced back to the presence of isolated
singularities in parameter space. Formally, Berry's phase can be
related to the curvature of the Hilbert space bundle over the base
space of parameters~\cite{Simon}, and it is this interpretation of
Berry's phase which renders its potential usefulness as a
diagnostic tool for criticality plausible: points of degeneracy
are associated with a greater curvature of the associated Hilbert
space bundle, and therefore it might be reasonable to expect
Berry's phase to act as a signature for quantum critical points in
interacting spin systems.

Of course, the geometric phase of $N$ non-interacting spin-$1/2$
particles with respect to a particular circuit is simply $N$ times
the geometric phase of a single particle, which in turn can be
related to the circuit's solid angle, as explained above. However,
as we switch on the spin-spin interactions, this simple
interpretation of the geometric phase in terms of solid angles
starts to break down. It is of great interest to know how the
geometric phase of a set of interacting spins with respect to a
given circuit changes as a function of the coupling parameter, and
whether the geometric phase is able to signal the presence of
critical points in the system. An important point to note here is
that the circuit in parameter space need only pass \emph{near} the
critical point for the Berry phase to register it. In other words,
the system need not actually undergo the quantum phase transition
for the Berry phase to pinpoint its presence and location in
parameter space, a consideration which assumes particular
importance in the light of the difficulties that may be associated
with physically implementing actual QPTs.

\subsection{Previous Work}
\label{subsection:PreviousWork}
A relationship between the geometric phase and criticality in spin
chains was examined by Carollo and
Pachos~\cite{Carollo1,Carollo2}. Specifically, the authors
analyzed the Berry phase of the ground state of a one-dimensional
spin-$1/2$ $XY$ model
\begin{equation}
  H = \sum_{l=-M}^{M} \left( \frac{1 + \gamma}{2}\sigma_l^x
      \sigma_{l+1}^x + \frac{1 - \gamma}{2}\sigma_l^y \sigma_{l+1}^y +
      \lambda \sigma_l^z \right),
\end{equation}
where $M = (N-1)/2$ for an odd number of spins $N$. The Berry
phase, as computed with respect to a rotation of the complete
Hamiltonian around the $z$-axis, was shown to exhibit the
following property in the thermodynamical limit
$N\rightarrow\infty$:
\begin{equation}
\label{eq:PT}
  \lim_{\gamma\rightarrow 0}(\varphi/M) =
    \begin{cases}
      0 \,(\text{mod }2\pi), & \text{for} |\lambda| > 1 \nonumber \\
      \text{finite}, & \text{for} |\lambda| < 1. \nonumber
    \end{cases}
\end{equation}
As the $XY$ model is critical for $|\lambda| < 1$ (and
non-critical otherwise), the quantity given above may be
interpreted as a signature of criticality for this model.
Unfortunately, the Berry phase \emph{per spin}, $\varphi/M$, is
not a physical quantity that is accessible via experiments. We can
only measure the \emph{total} Berry phase $\varphi$, which
vanishes for all values of $\lambda$: i.e.
$(1/M)\lim_{\gamma\rightarrow 0}\varphi = 0 \,(\text{mod } 2\pi),
\forall\lambda$. This is in agreement with Hamma~\cite{Hamma}, who
also showed, via a different line of argument, that for a finite
number of spins, the Berry phase acquired by the ground state is
always trivial ($0$ or $2\pi$). As an alternative, it was
suggested that the same quantity would be non-trivial in the
thermodynamic limit: $\lim_{\gamma\rightarrow 0}\lim_{M
\rightarrow \infty}(\varphi/M) \neq 0$. However, for any physical,
finite system, no matter how large, one is not able to detect
criticality in this manner. In other words, any finite-size
scaling behaviour is completely lacking, i.e. $\lim_{M \rightarrow
\infty}\lim_{\gamma\rightarrow 0}(\varphi/M) = 0\,\mbox{(mod
2$\pi$)}$; it is only in the thermodynamic limit itself that this
method represents a signature of criticality, i.e. the limiting
case is reached discontinuously. As such, the result's worth is
perhaps more of an abstract, mathematical nature, than of any
real, physically measurable consequence.

In the present work, in contrast, we employ a technique based on
the geometric phase that is able to detect critical points without
the need to contract the circuit's radius to zero or to extend the
number of particles to infinity. We do so by considering
discretized circuits that pass directly through the critical
point, rather than circumnavigating it. This procedure and the
obtained results are outlined in the following.

\section{Procedure and Outline}
\label{section:Outline}
We start by introducing the concept of a Bargmann
invariant~\cite{Bargmann} and its associated phase, which may be
regarded as a generalized Berry phase. It will be seen that
Bargmann invariants are ideally suited for \emph{numerically}
analyzing the gauge-invariant phase induced by evolving states
along a \emph{discretized} circuit in parameter space. We then
introduce the model of interacting spin-$1/2$ particles with which
we will be concerned for the rest of this paper and discuss some
of its more salient features. The Bargmann invariant and
associated phase is computed numerically for a chosen discretized
circuit and plotted as a function of the spin coupling strength
$J$. Of note will be the `speed' (to be defined) with which the
Bargmann invariant changes as a function of $J$ and the behaviour
of the magnitude of the Bargmann invariant; we will find that both
of these quantities can act as adequate signatures of criticality.
At this stage, it may be helpful to depict the procedure
schematically:
\begin{figure}[htp]
  \label{fig:schematic}
  \centering
  \psfrag{A}[tc][tc]{$J$}
  \psfrag{B}[tc][tc]{$J_c$}
  \psfrag{Sn}[bl][bl]{$\underline{\alpha}_{\mathcal{N}} = \underline{\alpha}_{0}$}
  \psfrag{SS}[rc][rc]{$\underline{\alpha}_{s+1}$}
  \psfrag{SS1}[rc][rc]{$\underline{\alpha}_{s}$}
  \includegraphics[width = \onefigurewidth]{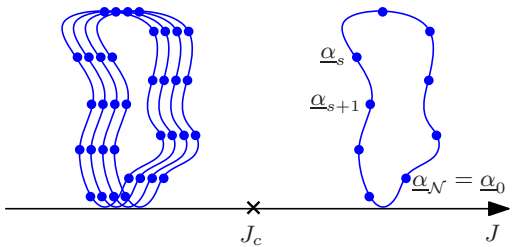}
  \caption{
A schematic representation of the procedure used to compute the
Bargmann invariant with respect to some chosen circuit
$\{\underline{\alpha}_s\}$ in parameter space, as a function of the
spin coupling strength $J$.
  }
\end{figure}
Let us fix the coupling parameter to some value $J_0$ and choose a
particular circuit composed of $\mathcal{N}$ vertices
$\underline{\alpha}_s$, each representing the set of parameters
needed (together with $J_0$) in order to completely describe the
Hamiltonian of the system at that point. At each individual point
on the circuit we numerically compute the ground state of the
system. The Bargmann invariant associated with the circuit at
$J_0$ is then obtained by cyclically multiplying successive
overlaps of these ground state wavefunctions. This will be
explained in more detail in section~\ref{section:Bargmann}, which
is devoted entirely to the subject of Bargmann invariants; for
now, we merely note that, in order to be sure to acquire a
gauge-invariant quantity in this way, the wavefunctions at the
start and end points of the circle need to be taken as
\emph{identical}. By repeating this procedure for a range of
different coupling strengths (generating a complex number, the
Bargmann invariant, for each value), we start to build up a
picture of the general trend. We present our results for varying
lengths of the spin chain, contrast the case of spin-$1/2$
particles with that of spin-$1$ particles, and discuss the effect
of increasing the number $\mathcal{N}$ of constituent vertices of
the circuit. Of particular interest, of course, is what, if
anything, may be said about the Bargmann invariant as the coupling
parameter crosses its critical point $J_c$. The report concludes
with a summary and brief discussion of our work.

\section{Bargmann Invariant and Phase}
\label{section:Bargmann}
In quite general terms one may define the phase difference $\chi$
between two non-orthogonal state vectors, $\ket{\Psi_1}$ and
$\ket{\Psi_2}$, by the relation
\begin{equation*}
    \exp{(i \chi)} \equiv
    \braket{\Psi_1}{\Psi_2}/|\braket{\Psi_1}{\Psi_2}|,
\end{equation*}
so that $\chi = \text{arg}\braket{\Psi_1}{\Psi_2}$. Naturally,
individual state vectors are only defined up to arbitrary phase
factors; $\chi$ is thus gauge-dependent and its value may not be
assigned any direct physical meaning. On the other hand, any
\emph{cyclic} combination of state vectors, such as
$\text{arg}(\braket{\Psi_1}{\Psi_2}\braket{\Psi_2}{\Psi_3}\braket{\Psi_3}{\Psi_1})$
is manifestly gauge-invariant and may therefore potentially
represent a physically relevant, measurable quantity. This
observation features strongly in the early works of
Pancharatnam~\cite{Pancharatnam} and it also appears in the form of
a remark in the celebrated proof of Wigner's theorem by
Bargmann~\cite{Bargmann}.

Introducing some notation, we define the \emph{Bargmann invariant}
with respect to a general $\mathcal{N}$-vertex circuit in parameter
space by
\begin{equation}
  \label{eq:phasedefn2}
  \mathcal{C} =
  \prod_{s=0}^{\mathcal{N}-1}\braket{\psi_s}{\psi_{s+1}} =
  \tr{\prod_{s=0}^{\mathcal{N}-1}\proj{\psi_s}},
  \quad
  \psi_{\mathcal{N}} = \psi_{0}.
\end{equation}
The associated \emph{Bargmann phase} is denoted by $\varphi =
\text{arg}(\mathcal{C})$. Modulo $2\pi$, the global Bargmann phase
is just the sum of the individual phases, i.e. $\varphi =
\sum_{s=0}^{\mathcal{N}-1}\text{arg}\braket{\psi_s}{\psi_{s+1}}$.
Note that we require $\mathcal{N} \geqslant 3$ for non-trivial
Bargmann phases.

Suppose now, for argument's sake, that we are dealing with state
vectors $\ket{\psi(\underline{\alpha}_s)}$ that represent the
instantaneous ground states of a family of parameter-dependent
Hamiltonians $H(\{\underline{\alpha}\})$. Then, in the continuum
limit where the sum above is replaced by a contour integral over an
infinitesimal phase along a (smooth) closed circuit in parameter
space, the Bargmann invariant reduces to the usual Berry phase. In
this sense Bargmann invariants may be regarded as generalized Berry
phases. The key advantage afforded by the generalized formulation
lies in its computational ease: Bargmann invariants lend themselves
directly to numerical computation, and thus, crucially, to scenarios
where parts of the circuit under consideration represent
non-integrable Hamiltonians, as is the case in the present work. An
obvious potential drawback is that the discretization procedure may
lack the simple interpretational underpinnings of the Berry phase in
terms of physical adiabatic processes.

\section{The Spin Model}
\label{section:SpinModel}
Consider a spin chain with nearest-neighbour $XX$ interactions in
the presence of an external magnetic field, which is aligned along
an arbitrary direction $\overrightarrow{n}$. The system is
described by the following Hamiltonian:
\begin{equation}
    \label{eq:generalham}
    H_{\overrightarrow{n}} = J \sum_{k=1}^{N}\sigma^{x}_{k}\sigma^{x}_{k+1}
            + B \sum_{k=1}^{N}\sigma^{\overrightarrow{n}}_{k},
            \quad N \geqslant 2,
\end{equation}
where, $\overrightarrow{n} = (n_x,n_y,n_z)$ denotes a vector of
unit length, so that $\sigma^{\overrightarrow{n}} = \sum_{\alpha}
n_{\alpha}\sigma^{\alpha}$ with $\sum_{\alpha} n_{\alpha}^2 = 1$,
$\alpha \in \{x,y,z\}$. Note that we impose periodic boundary
conditions, so that $\sigma^{x}_{N+1} \equiv \sigma^{x}_1$.

As already outlined in the previous two sections, our aim is now
to analyze the Bargmann invariant induced by the ground states of
a family of Hamiltonians (\ref{eq:generalham}) that is
characterized by a set of $\mathcal{N}$ unit vectors
$\{\overrightarrow{n}\}$. Typically, we will imagine the magnetic
field vectors of that family of Hamiltonians to be arranged on the
vertices of a regular polygon residing on the surface of the Bloch
sphere (i.e. $B = 1$); an example of such a circuit is depicted in
Fig.~\ref{fig:Bloch2}. For a given coupling strength, $J_0$, the
Bargmann invariant can now be obtained by numerically computing
the ground states at each of the circuit's vertices and applying
Eq.~(\ref{eq:phasedefn2}). Having chosen our circuit in parameter
space, we would then like to investigate the manner in which the
Bargmann invariant depends on the choice of spin coupling
strength, $J$. Of particular interest are circuits that slice
through any critical points of the system. Does the Bargmann
invariant undergo a marked shift when the parameters pass through
their critical values? Of further interest is the dependence on
the number $\mathcal{N}$ of vertices of the regular polygon
circuit, as well as the dependence on the number $N$ of spins in
the chain.

Leaving the choice of circuit in parameter space open for now, it
is evident from the form of Hamiltonian (\ref{eq:generalham}) that
there are two `special' directions in which the magnetic field
vector may point, namely the $x$-direction and any direction in
the $y-z$ plane. The former is essentially a classical model while
the latter corresponds to the quantum Ising model. In both cases,
the Hamiltonian is analytically soluble. Of course, it is well
known that the Ising model exhibits criticality when $B = J$,
provided we find ourselves in the thermodynamic limit of an
infinite spin chain. Aside from the two special cases outlined,
however, the Hamiltonian will in general be non-integrable at any
point along the circuit; this forces us to resort to numerical
simulations of a limited number of spins, which moves the Ising
model's critical point outside our region of accessibility. On the
other hand, when the magnetic field vector points in the
$x$-direction, the Hamiltonian does possess a critical point, even
for a \emph{finite} number of spins, as outlined in the following.

\section{The Critical Point} \label{section:CriticalPoint}
A special case of the class of Hamiltonians (\ref{eq:generalham})
occurs when the magnetic field vector points in the $x$-direction:
\begin{equation}
  \label{eq:specialham}
    H_x =
    J \sum_{k=1}^{N}\sigma^{x}_{k}\sigma^{x}_{k+1}
    + B \sum_{k=1}^{N}\sigma^{x}_{k}, \quad N \geqslant 2.
\end{equation}
This Hamiltonian is purely classical: its two constituent terms
commute and are diagonal in the \{\ket{\pm}\} eigenbasis. In the
following we will demonstrate that $H_x$ possesses a critical
point~\cite{footnote1} at $J=J_c \equiv |B|/2$.

It is not difficult to find the ground state of this model. When
$J \leqslant J_c$, it is simply
\begin{equation}
\label{eq:PT}
  \ket{E^{(0)}} =
    \begin{cases}
      \ket{-}^{\otimes N}, \text{  for }B \geqslant 0 \, \& \, J \leqslant J_c,& \nonumber \\
      \ket{+}^{\otimes N}, \text{  for }B \leqslant 0 \, \& \, J \leqslant J_c,& \nonumber
    \end{cases}
\end{equation}
with the corresponding (non-degenerate) ground state energy given
by $E^{(0)} = N(J-2J_c)$.

In the regime $J \geqslant J_c$, the situation is only slightly
more involved as the parity of $N$ now becomes important. For an
\emph{even} number of spins, the (doubly degenerate) ground state
energy is given by $E^{(0)} = -NJ$, the corresponding eigenstates
being $\ket{+-}^{\otimes N/2}$ and $\ket{-+}^{\otimes N/2}$. For
an \emph{odd} number of spins, the ground state energy is $N$-fold
degenerate and reads $E^{(0)} = -NJ + 2(J-J_c)$. The ground state
is given by
\begin{equation}
\label{eq:PT}
  \ket{E^{(0)}} =
    \begin{cases}
      \ket{-+-+- \cdots -+-}, \text{  for }B \geqslant 0 \, \& \, J \geqslant J_c,& \nonumber \\
      \ket{+-+-+ \cdots +-+}, \text{  for }B \leqslant 0 \, \& \, J \geqslant J_c,& \nonumber
    \end{cases}
\end{equation}
and all $N$ translations thereof. In order to convince oneself
that this is indeed the correct ground state, one can easily show
that no individual spin flip is capable of lowering the energy any
further.

From the analysis above it follows that the ground state energy
$E^{(0)}$ can always be written as a function of $|J-J_c|$,
demonstrating non-analyticity at the critical point $J_c$.
Moreover, the energy gap vanishes in the region $J \geqslant J_c$.

\section{The Circuit} \label{section:CriticalPoint}
\begin{figure}[htp]
  \centering
  \includegraphics[width = .348\textwidth]{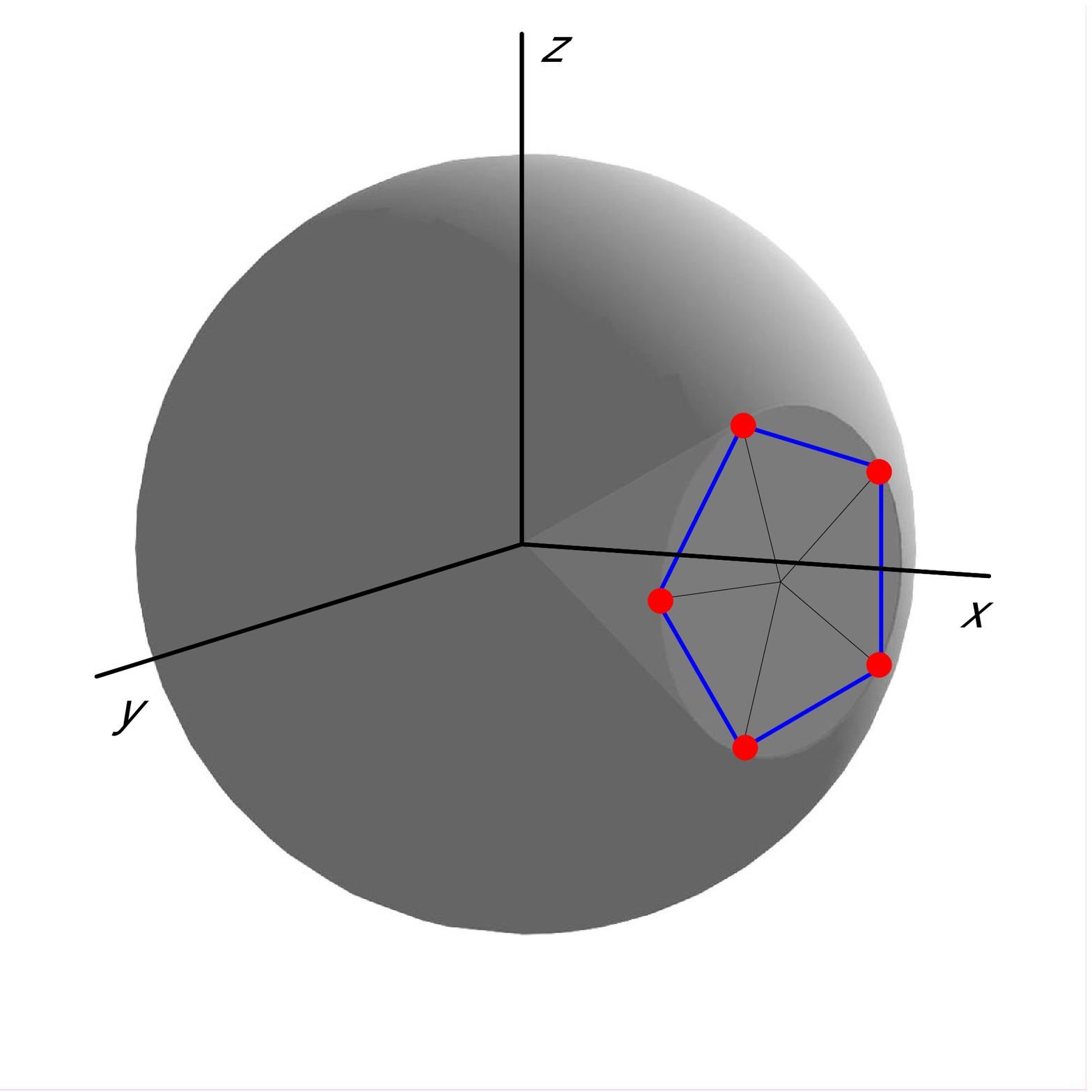}
  \caption{
A schematic representation of the circuit's topology. The critical
$x$-axis slices mid-way through an edge that connects two adjacent
vertices of the polygon.
  }
  \label{fig:Bloch2}
\end{figure}

In light of the previous discussion, the parameter circuit which
suggests itself is the following: a regular polygon residing on
the Bloch sphere of which one edge is bisected by the $x$-axis, as
illustrated in Fig.~\ref{fig:Bloch2}.
As we move along the vertices of the circuit, the critical $x$-axis
is crossed in the direction of positive $z$-axis.

\section{Results for the Spin-$1/2$ Chain}
\label{section:Results}
\begin{figure}[htp]
  \centering
%  \psfrag{Xlabel}[tc][tc]{$(J-J_c)/\text{radius}$}%
%  \psfrag{Ylabel}[bc][bc]{$\varphi/\pi$}%
%  \psfrag{replaceX}[tc][tc]{$(J-J_c)/\text{radius}$}%
%  \psfrag{replaceY}[bc][bc]{$\varphi/\pi$}%
  \psfrag{Xlabel}[tc][tc]{}%
  \psfrag{Ylabel}[bc][bc]{}%
  \psfrag{replaceX}[tc][tc]{}%
  \psfrag{replaceY}[bc][bc]{}%
  \hspace{-0.1cm}%
  \subfigure{\label{subfig:3cnum}
    \includegraphics[width = .22\textwidth]{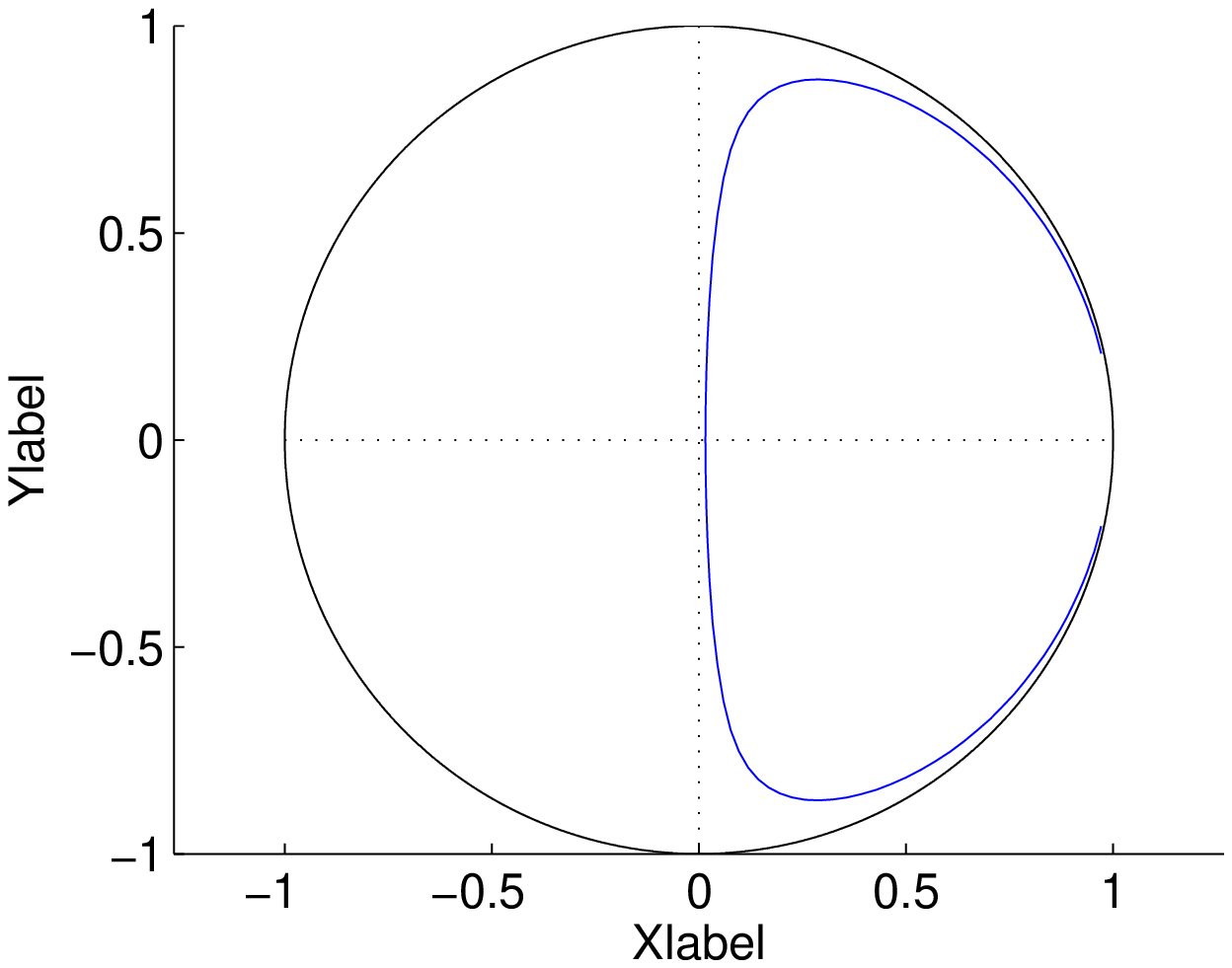}
  }\hspace{0.2cm}%
  \subfigure{\label{subfig:3phase}
    \includegraphics[width=.1815\textwidth,height=.1727\textwidth]
      {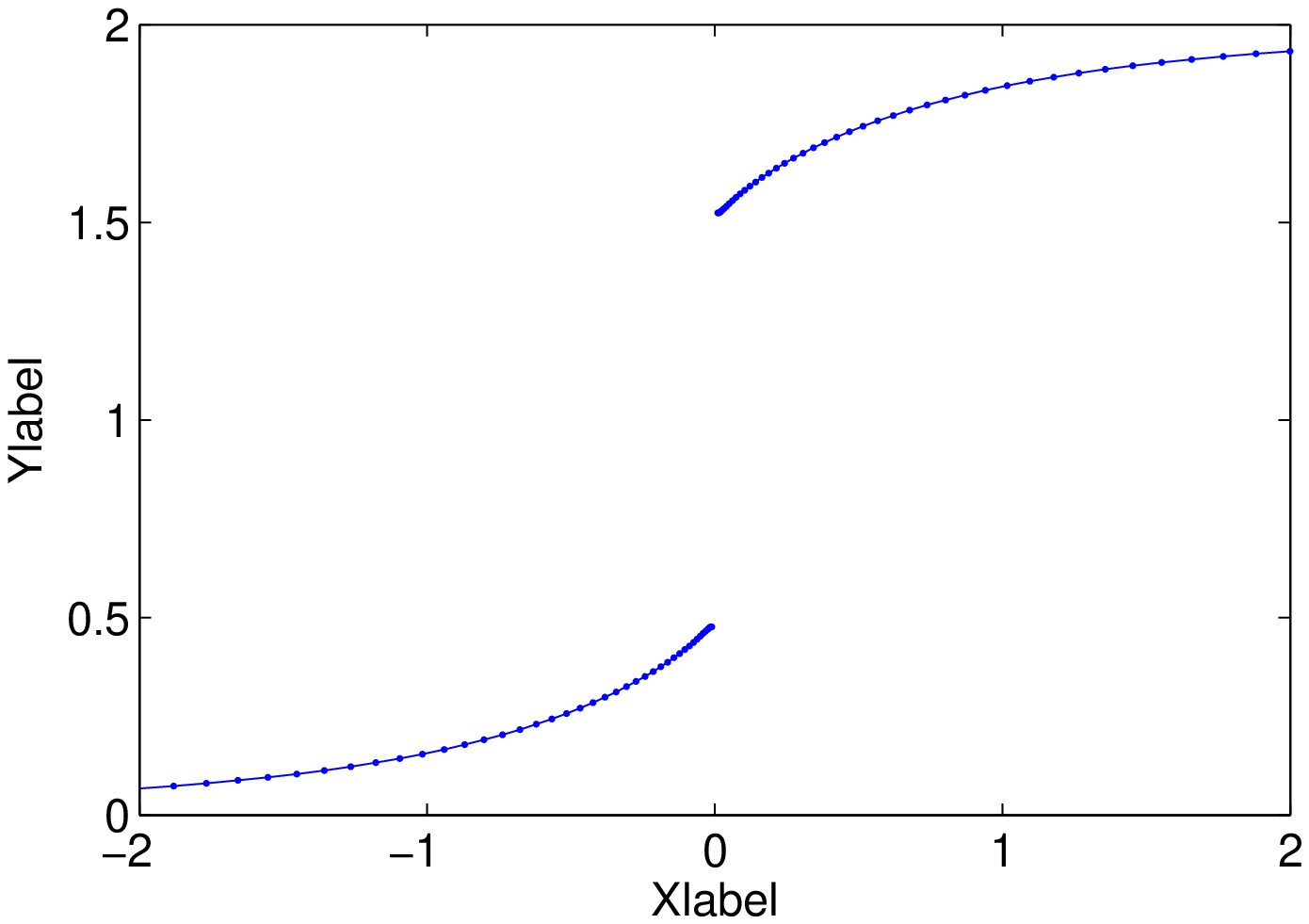}
  }\\[-0.2cm]%
  \hspace{-0.1cm}%
  \subfigure{\label{subfig:5cnum}
    \includegraphics[width = .22\textwidth]{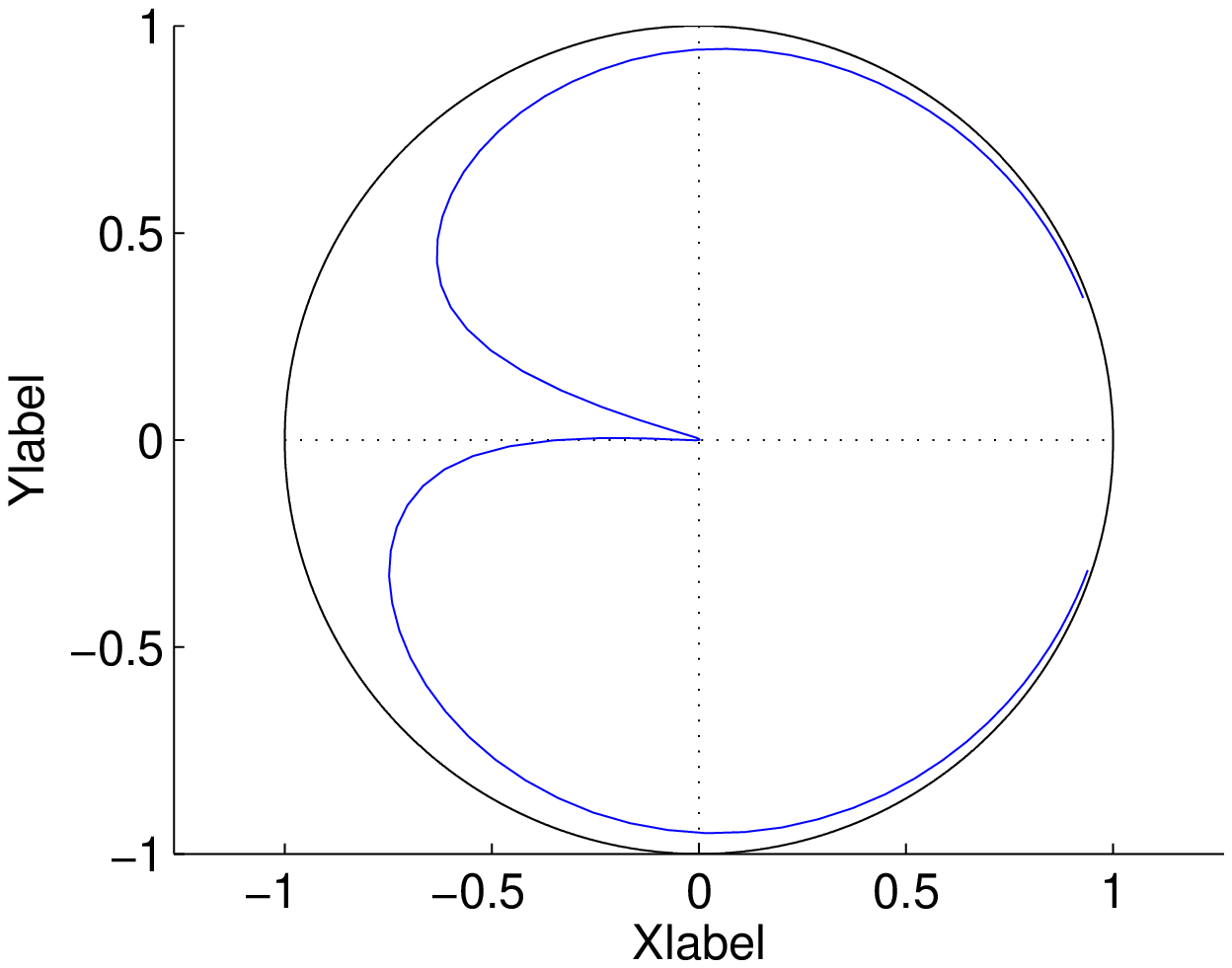}
  }\hspace{0.2cm}%
  \subfigure{\label{subfig:5phase}
    \includegraphics[width=.1815\textwidth,height=.1727\textwidth]
      {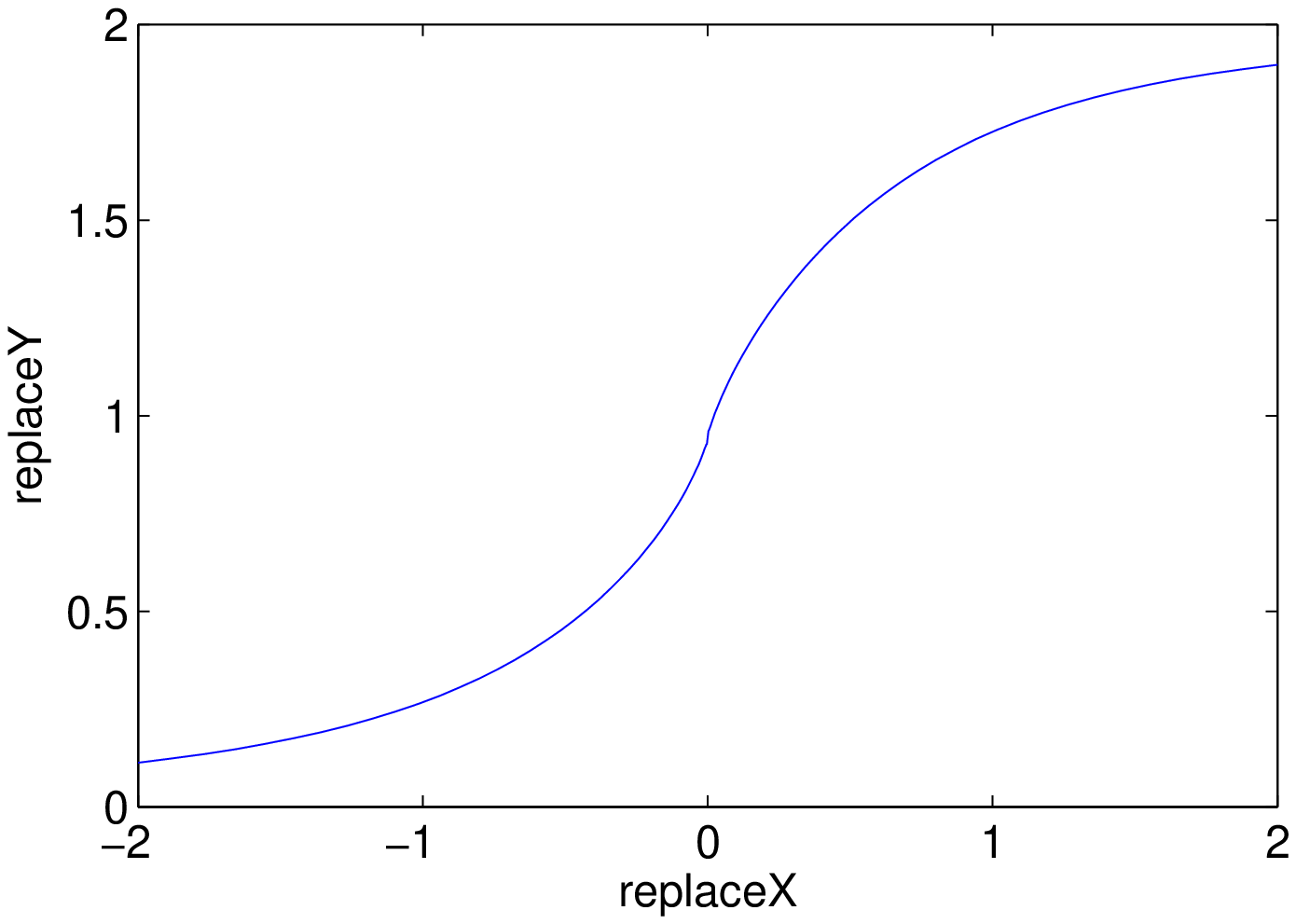}
  }\\[-0.2cm]%
  \hspace{-0.1cm}%
  \subfigure{\label{subfig:7cnum}
    \includegraphics[width = .22\textwidth]{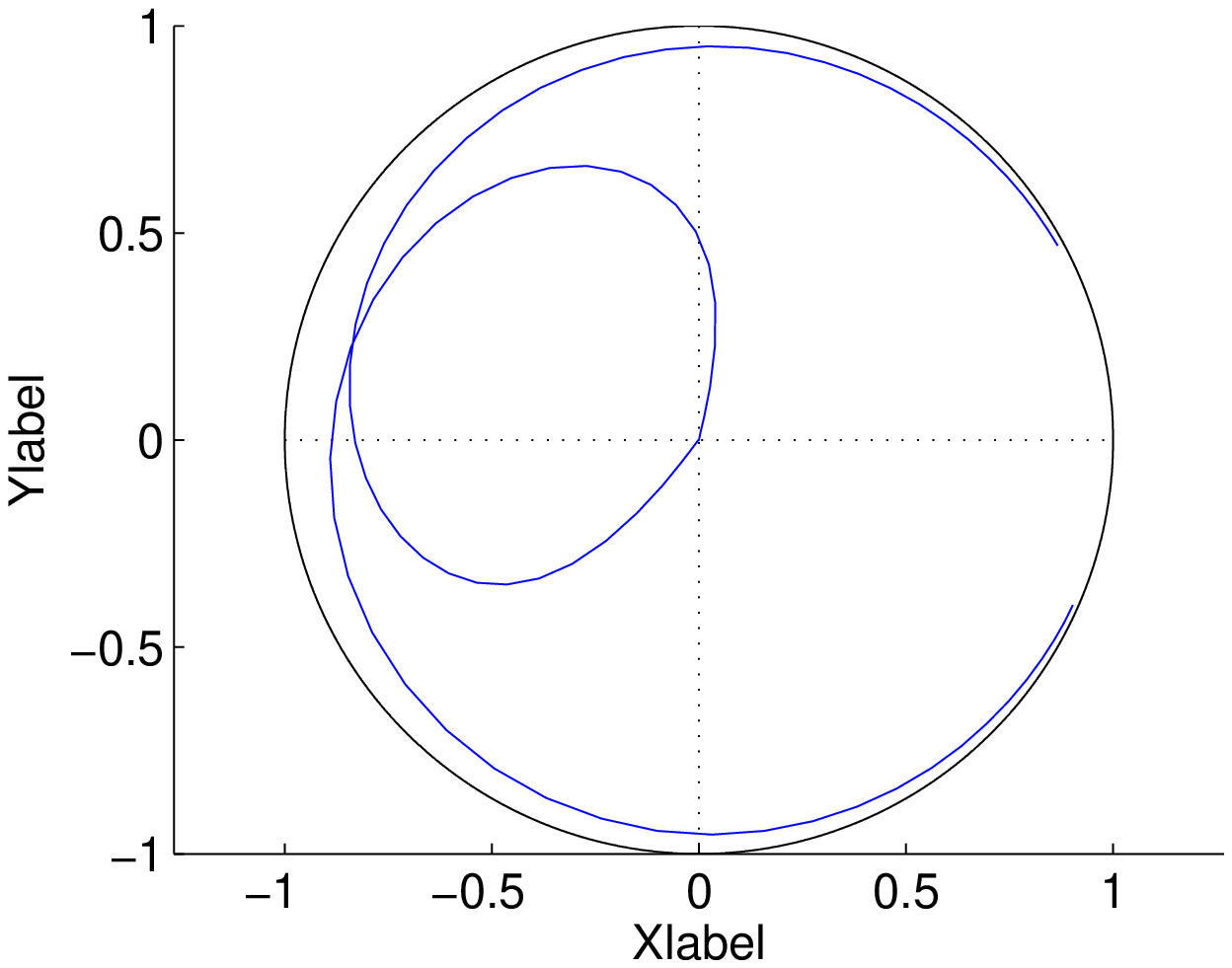}
  }\hspace{0.2cm}%
  \subfigure{\label{subfig:7phase}
    \includegraphics[width=.1815\textwidth,height=.1727\textwidth]
      {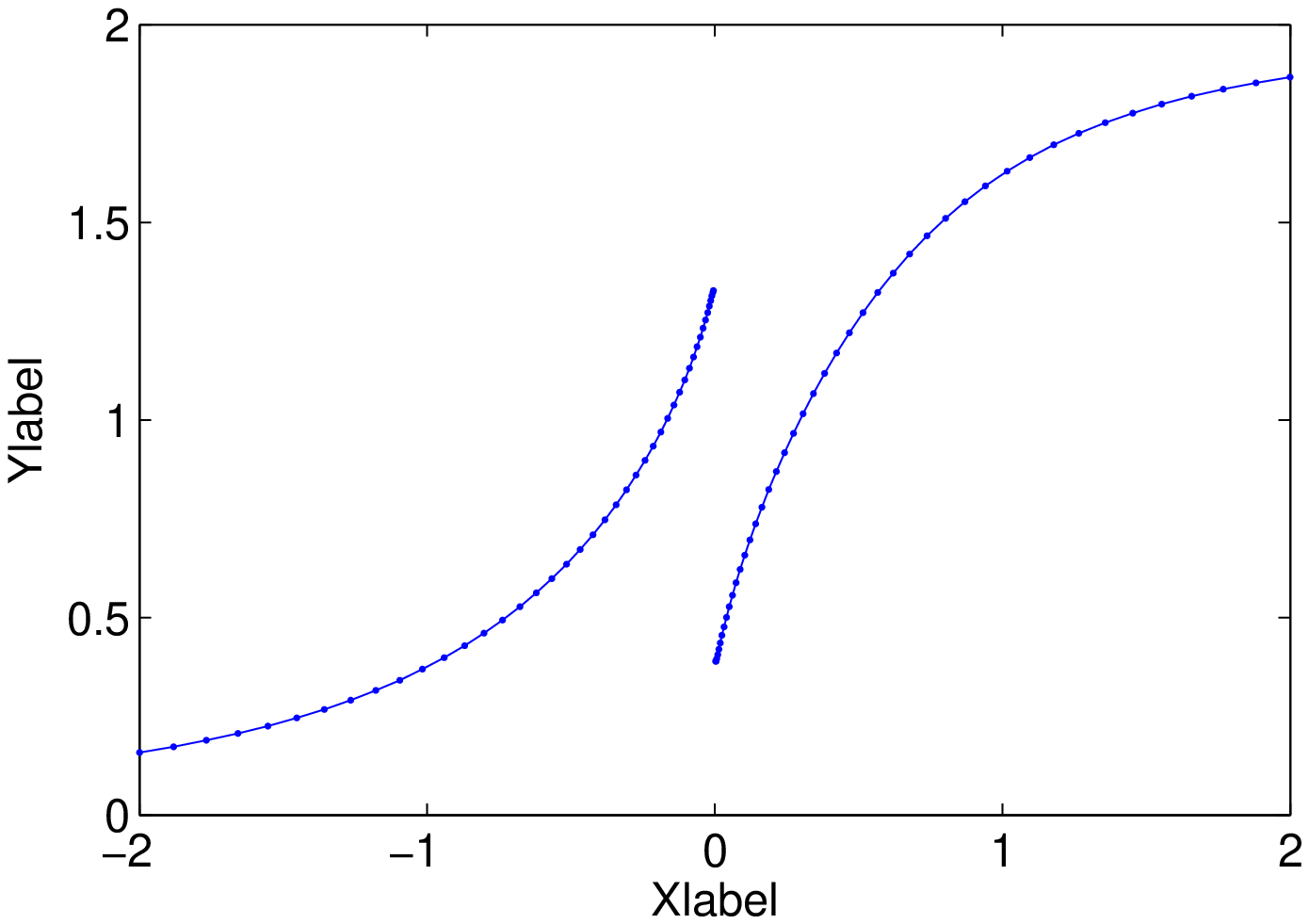}
  }\\[-0.2cm]%
  \hspace{-0.1cm}%
  \subfigure{\label{subfig:9cnum}
    \includegraphics[width = .22\textwidth]{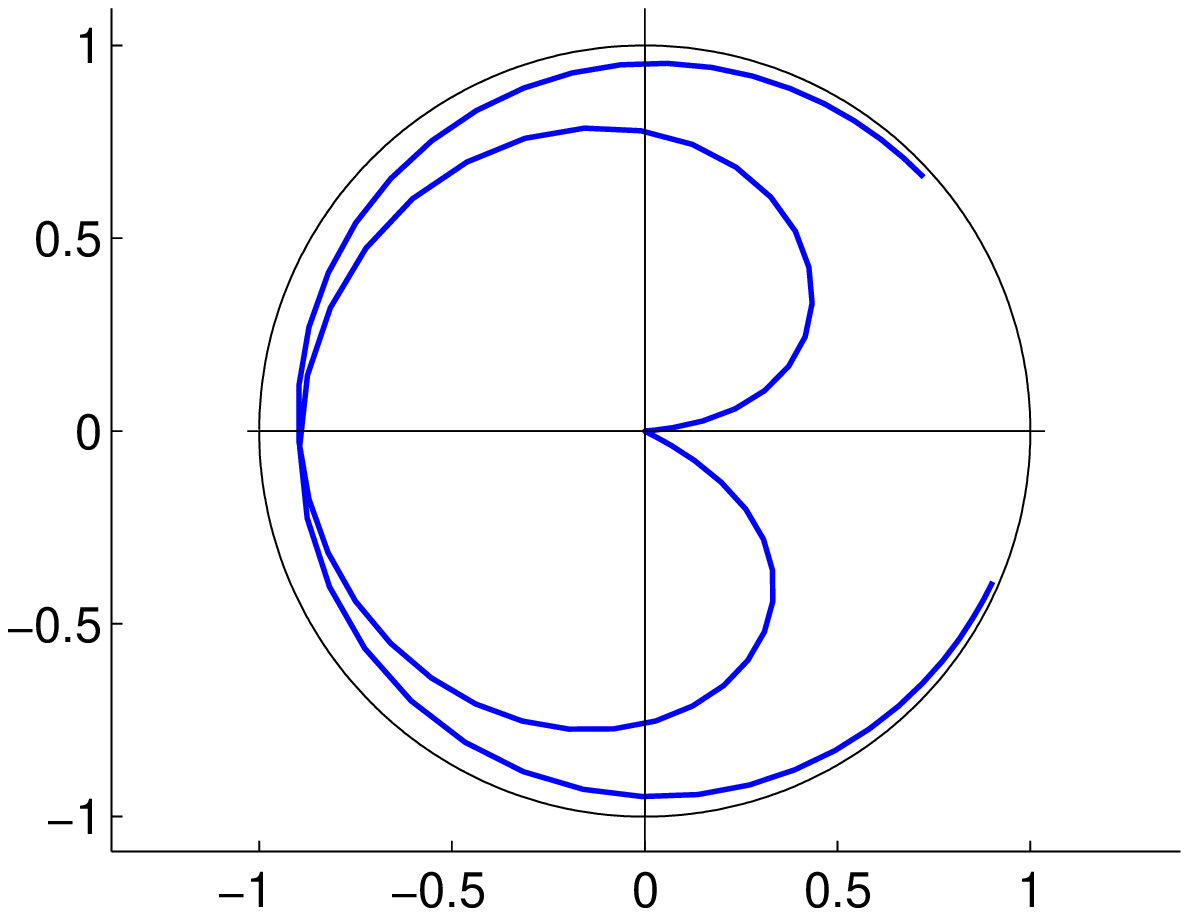}
  }\hspace{0.2cm}%
  \subfigure{\label{subfig:9phase}
    \includegraphics[width=.1815\textwidth,height=.1727\textwidth]
      {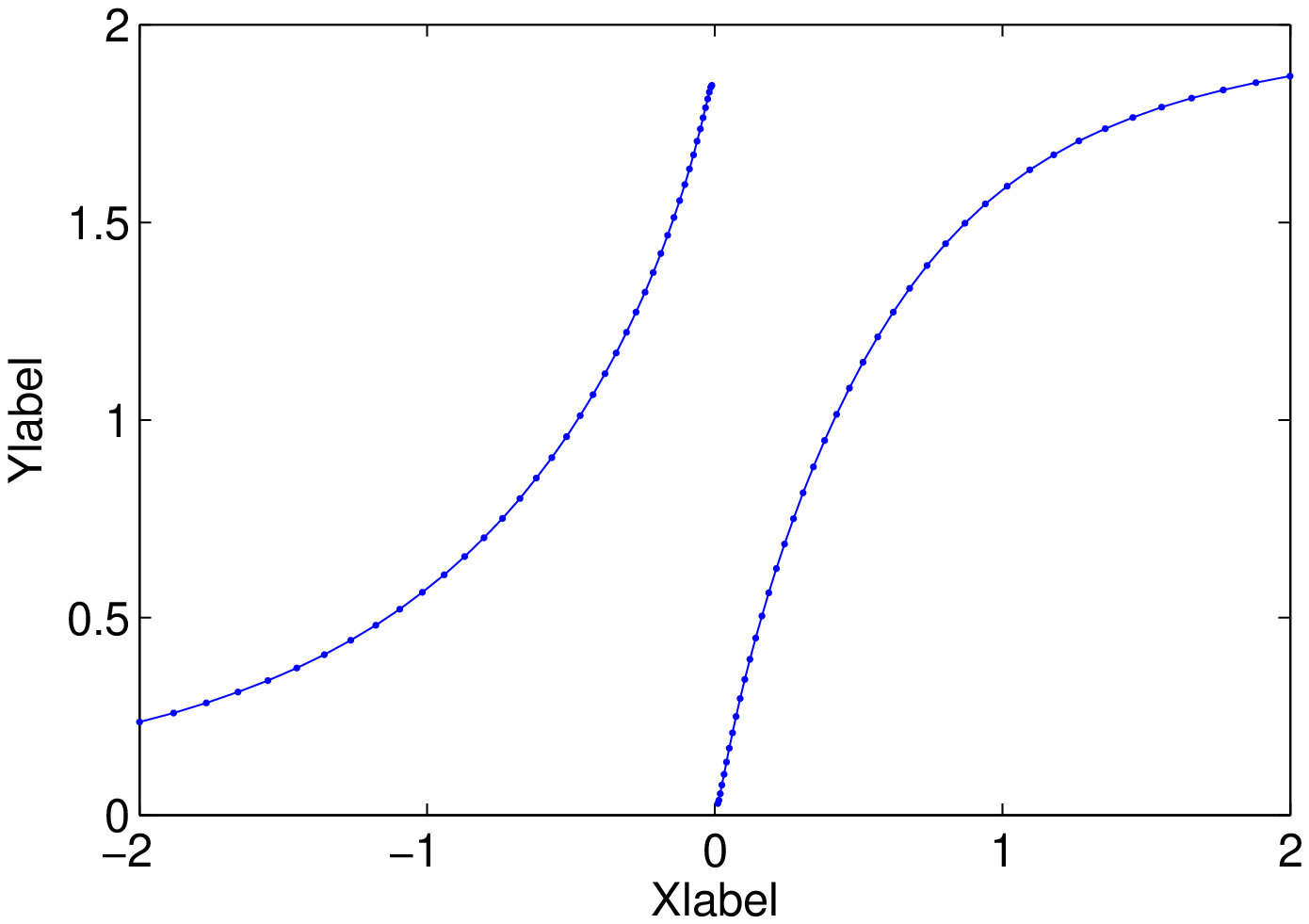}
  }\\[-0.2cm]%
  \hspace{-0.1cm}%
  \subfigure{\label{subfig:11cnum}
    \includegraphics[width = .22\textwidth]{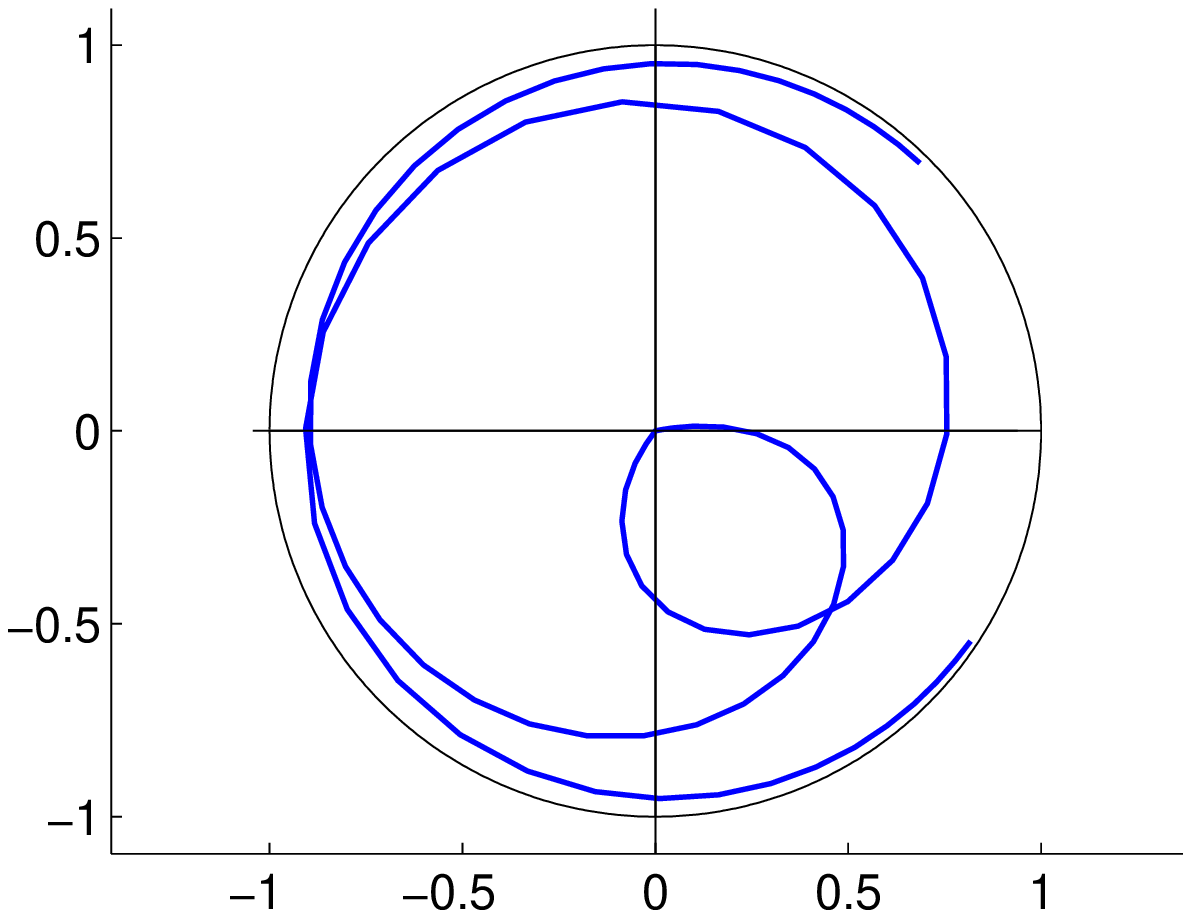}
  }\hspace{0.2cm}%
  \subfigure{\label{subfig:11phase}
    \includegraphics[width=.1815\textwidth,height=.1727\textwidth]
      {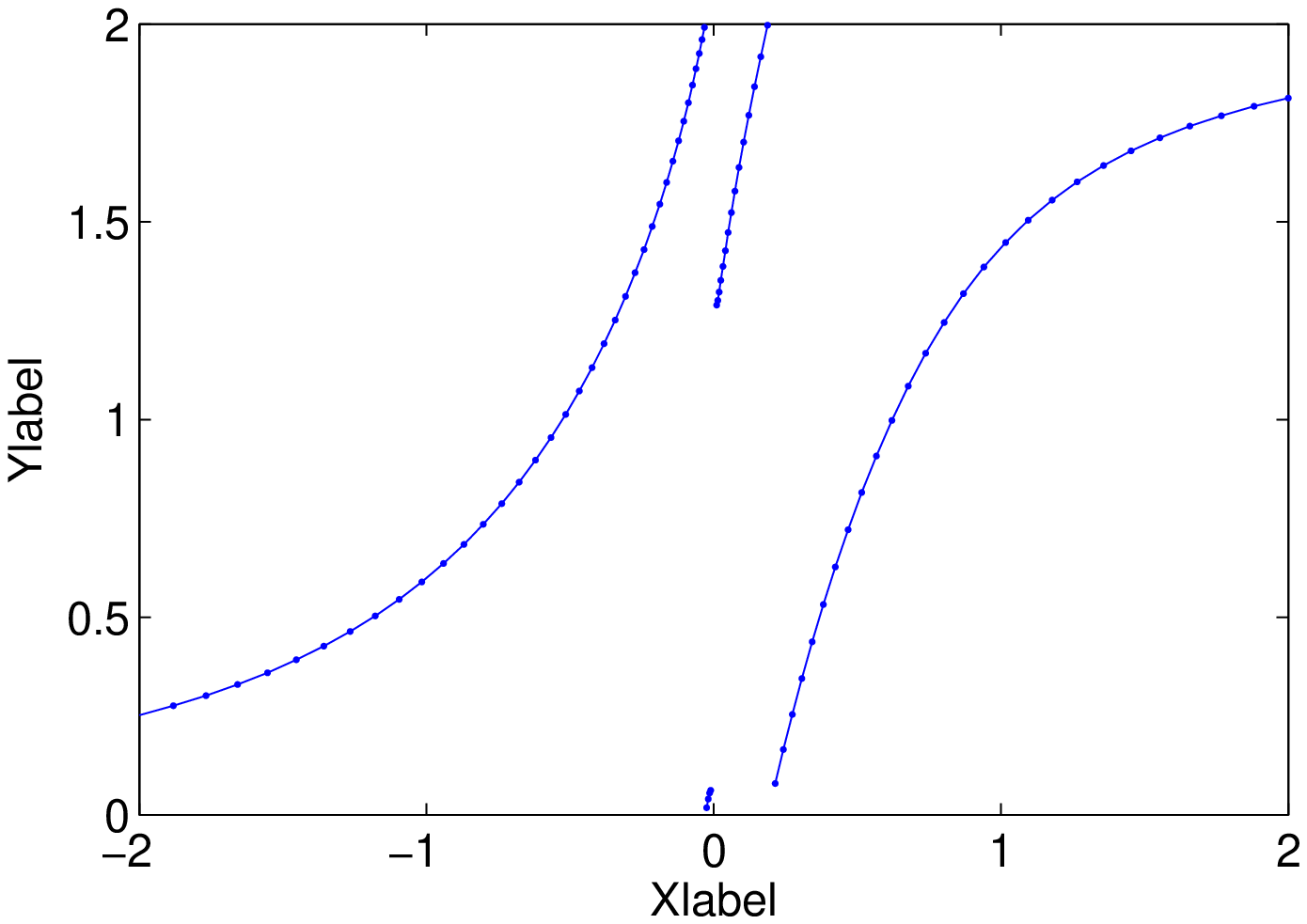}
  }\\[-0.2cm]%
  \caption{
Bargmann invariants and their phases. The rows correspond to the
number of spin-$\frac{1}{2}$ particles $N$, increasing from the top
as $3,5,7,9,11$. The left column depicts the Bargmann invariants
$\mathcal{C}$ on the unit disk in the complex plane, each point of
the graph stemming from a different value of the coupling parameter
$J$. The right-hand column shows the corresponding Bargmann phase
$\varphi/\pi$ as a function of the coupling parameter $(J-J_c)/r$.
See the text for a detailed description of observations.
  }\label{fig:spinonehalf}
\end{figure}

We now turn to the results of our simulation of the Bargmann
invariant for the spin-$1/2$ chain, which are summarized in
Fig.~\ref{fig:spinonehalf}. The circuit used in the simulation is
the regular polygon described in the previous section, but
consists of $\mathcal{N}=100$ vertices.
The Bargmann invariant and its phase were computed for different
spin chain lengths, ranging in odd numbers from $3$ to $11$, and
shown in Fig.~\ref{fig:spinonehalf} as increasing from the top.
Note that the results for even numbers of spins have been omitted
as they showed no qualitative difference. This may be due to
finite-size effects of the kind which have been observed in
similar studies elsewhere, see for example Ref.~\cite{Kay}. The
left-hand column of the figure depicts the Bargmann invariants
$\mathcal{C}$ on the unit disk in the complex plane. Each data
point of the graph stems from a different value of the coupling
parameter $J$. The right-hand column shows the corresponding
Bargmann phase $\varphi/\pi$, as a function of the coupling
parameter $(J-J_c)/r$, where $r$ refers to the radius of the
circle within which the $\mathcal{N}$-polygon circuit is
inscribed. Note that in this case the radius was chosen as
$10^{-5}$, but the graphs, as they are plotted, are qualitatively
independent of the size of this radius.

We would like to point out several features that are apparent from
Fig.~\ref{fig:spinonehalf}. First and foremost, the Bargmann
invariant $\mathcal{C}$ is observed to always pass through the
origin of the complex plane just when the spin coupling strength
attains its critical value of $J_c = 1/2$. In other words, the
magnitude $|\mathcal{C}|$ vanishes at the critical point. This is
one of the features which we would like to propose as a signature
of criticality, and is discussed further in
section~\ref{subsection:magsignature}.

Another trend that is immediately apparent from
Fig.~\ref{fig:spinonehalf} is that with increasing spin chain
length the Bargamann invariants `wind ever more loops' around the
origin. This statement can be made a little more precise by
referring to the corresponding Bargmann phases. These graphs are
characterized by discontinuities at the critical point: `jumps' by
$\pi$ alternating with `jumps' by $2\pi$. However, on joining up
those separate pieces in the graphs end-to-end, one notices that
the `image' covered by the resulting graphs grows with $N$ as $\pi
(N-1)/2$. This result may be explained by way of referring to the
$J=0$ limit of non-interacting spins for which the Berry phase
grows linearly with the number of spins. Also note that for
spin-$1$ chains, the `joined-up' Bargmann phase covers twice as
much ground for any given number of particles (see
Fig.~\ref{fig:spinone}).

In addition, we should note that we obtain Berry's solid angle
result for the case of non-interacting spins, $J=0$. Approximating
the solid angle of our regular polygon circuit as that of the
enveloping cone, the Bargmann phase $\varphi$ asymptotically
approaches Berry's solid angle result as the number of vertices of
the polygon is increased.

Of course, in the present work only spin chains of up to $11$
particles were simulated, and the exponential growth in the size of
the Hilbert space with increasing particle number precludes one from
simulating chains that are much longer than that~\cite{footnote2}.
However, from the preceding discussion the trend for Bargmann
invariants and phases of longer spin chains should by now be fairly
obvious. In particular, it has become clear that, perhaps
surprisingly, precious little may be deduced about criticality from
the Bargmann phase. Instead, a much clearer indication of the
existence of a critical point is the vanishing magnitude of the
Bargmann invariant itself.

\section{Signatures of Criticality}
\subsection{Magnitude of the Bargmann Invariant}
\label{subsection:magsignature}
\begin{figure}[htp]
  \centering
  \psfrag{Xlabel}[tc][tc]{$(J-J_c)/r$}
  \psfrag{Ylabel}[bc][bc]{$|\mathcal{C}|$}
  \includegraphics[width=.45\textwidth, height=.3\textwidth]{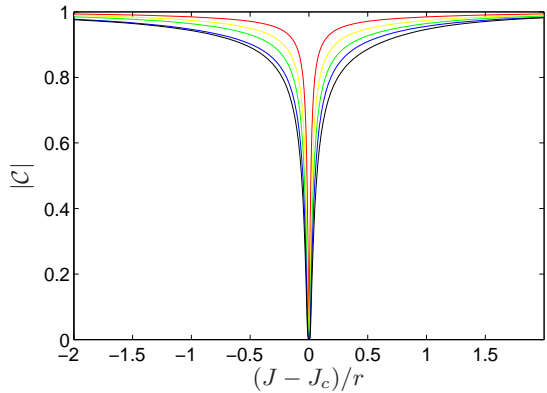}
  \caption{
Plots of $|\mathcal{C}|$ for varying lengths of the spin chain,
ranging from $N=3$ to $N=11$ (odd numbers only). The longer the
spin chain, the more sharply peaked is the graph. The number of
vertices on the circuit is kept constant, at $\mathcal{N} = 100$.
  }\label{fig:varyspinnumber}
\end{figure}

\begin{figure}[htp]
  \centering
  \psfrag{Xlabel}[tc][tc]{$(J-J_c)/r$}
  \psfrag{Ylabel}[bc][bc]{$|\mathcal{C}|$}
  \psfrag{a}[tl][tl]{$\mathcal{N} = 300$}
  \psfrag{b}[tl][tl]{$\mathcal{N} = 100$}
  \includegraphics[width=.45\textwidth, height=.3\textwidth]{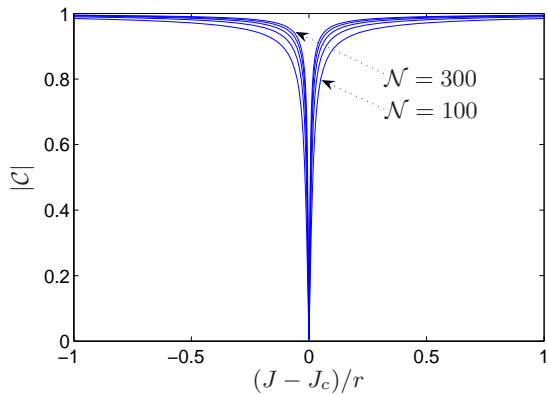}
  \caption{
Plots of $|\mathcal{C}|$ for varying numbers of vertices on the
circuit, ranging from $\mathcal{N}=100$ to $\mathcal{N}=300$ (in
steps of $50$). The more vertices there are on the circuit, the
more sharply peaked is the graph. The spin chain length is kept
constant, at just $N = 3$. Note that qualitatively nothing changes
as we increase the chain length (tested up to $N=13$).
  }\label{fig:varypointnumber}
\end{figure}

It was already apparent from Fig.~\ref{fig:spinonehalf} that the
magnitude of the Bargmann invariant $\mathcal{C}$ is able to
assume the role of a diagnostic tool for critical points in the
finite-sized interacting spin chain considered. For clarity, the
magnitude of the graphs of $\mathcal{C}$ in
Fig.~\ref{fig:spinonehalf} are plotted separately in
Fig.~\ref{fig:varyspinnumber}.
An analogous plot is shown in Fig.~\ref{fig:varypointnumber}, but
this time for a fixed number of spins $N$ and a varying number of
circuit vertices $\mathcal{N}$.
Of course, there is a perfectly plausible explanation for these
results: as the circuit passes over the critical point the ground
state changes abruptly, so that the overlap of the ground states
on either side of the critical point will be rather small in
magnitude. This in turn forces the magnitude of the product of
overlaps along the circuit in Eq.~\ref{eq:phasedefn2} to be small.
In fact, similar results were observed in a work by Zanardi and
Paunkovi$\acute{\text{c}}$~\cite{Zanardi}, who concluded that the
ground state overlap function is itself a good characterization of
QPTs.

\subsection{`Speed' of the Bargmann Invariant}
\begin{figure}[htp]
  \centering
  \psfrag{Xlabel}[tc][tc]{$(J-J_c)/r$}
  \psfrag{Ylabel}[bc][bc]{\text{`speed'} $v$}
  \includegraphics[width=.45\textwidth, height=.3\textwidth]{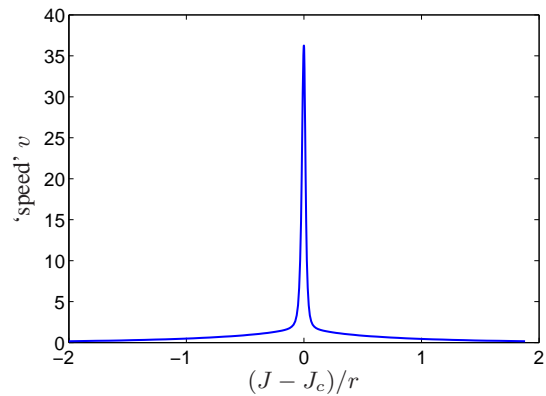}
  \caption{
Plot of the `speed', $v_s :=
|\mathcal{C}_{s+1}-\mathcal{C}_s|/(J_{s+1} - J_s)$, as a function
of the spin coupling strength. The spin chain length is $N=3$ and
the circuit is composed of $\mathcal{N} = 100$ vertices.
  }\label{fig:speed}
\end{figure}

Another characteristic that is sensitive to the phase transition
is the `speed' with which the Bargmann invariant changes as a
function of the coupling parameter, which we define by $v_s :=
|\mathcal{C}_{s+1}-\mathcal{C}_s|/(J_{s+1} - J_s)$. As is evident
from Fig.~\ref{fig:speed}, the speed picks up markedly in the
vicinity of the critical point. Again, this phenomenon is a direct
result of the rapid change of the ground state near a critical
point.

\section{Results for the Spin-$1$ Chain}
\label{section:spinone}
\begin{figure}[htb]
  \centering
  \subfigure{\label{subfig:5cnum}%
    \includegraphics[width = \figurewidth]{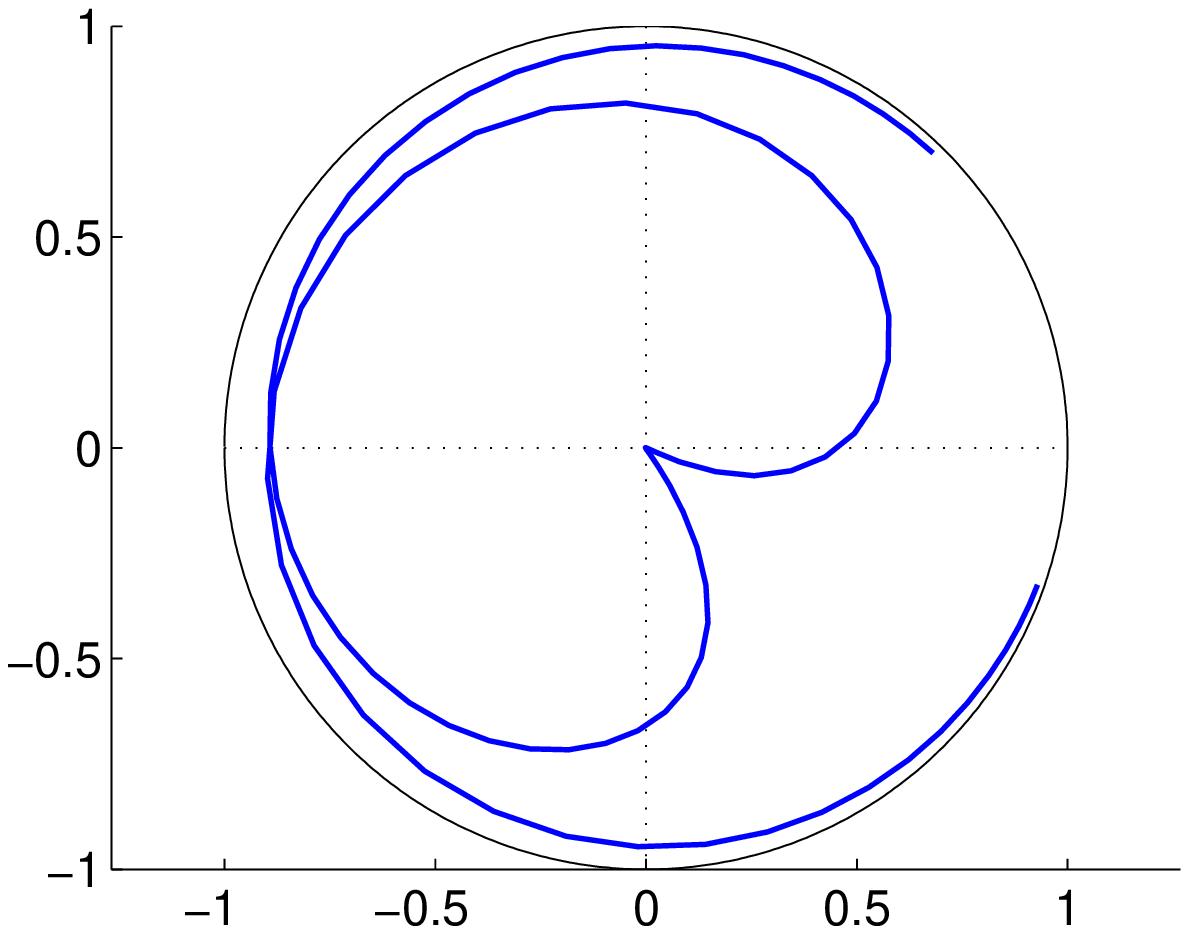}
  }\hspace{.25cm}%
  \subfigure{\label{subfig:7cnum}
    \includegraphics[width = \figurewidth]{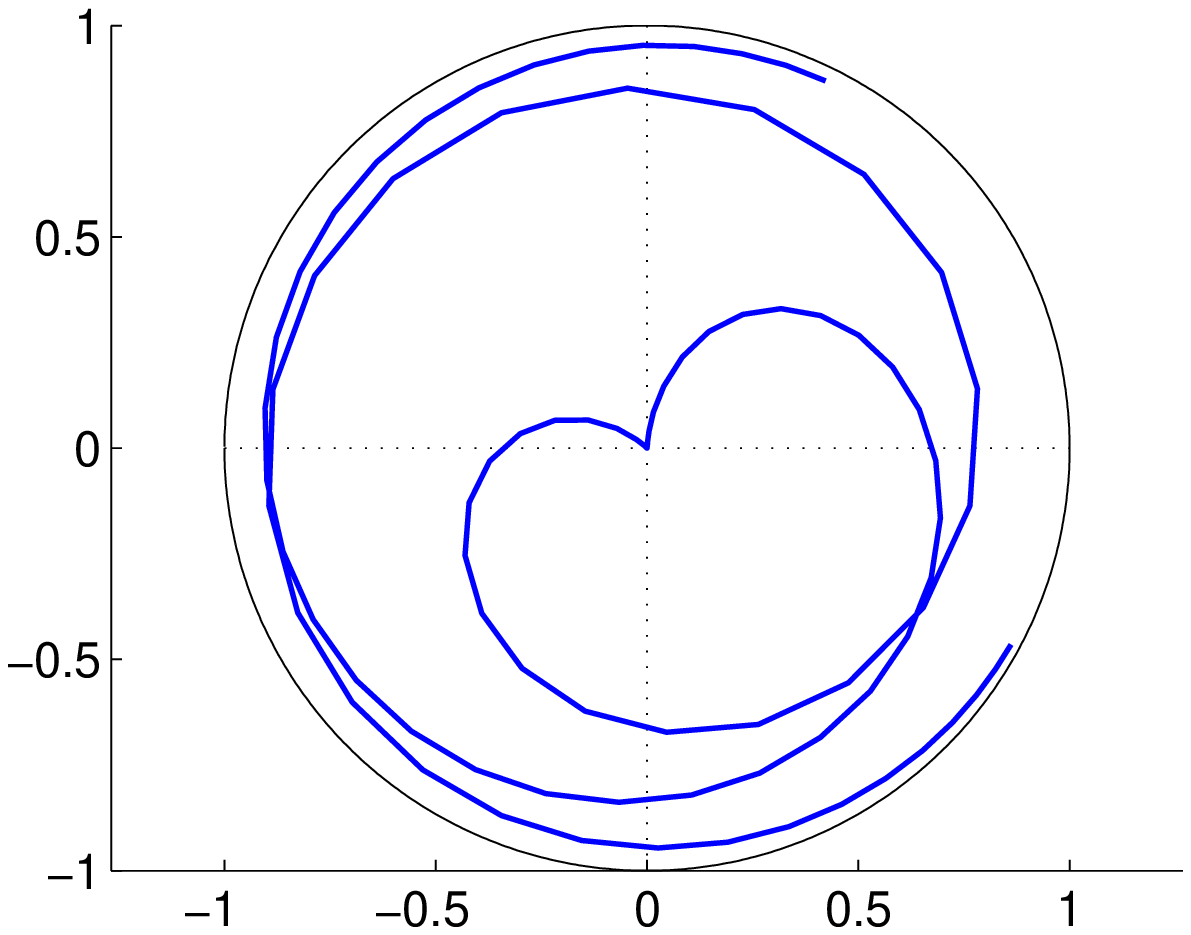}
  }\\[0.2cm]%
  \psfrag{Xlabel}[tc][tc]{$(J-J_c)/r$}
  \psfrag{Ylabel}[bc][bc]{$\varphi/\pi$}
  \subfigure{\label{subfig:5phase}
    \includegraphics[width=.1815\textwidth,height=.1727\textwidth]
      {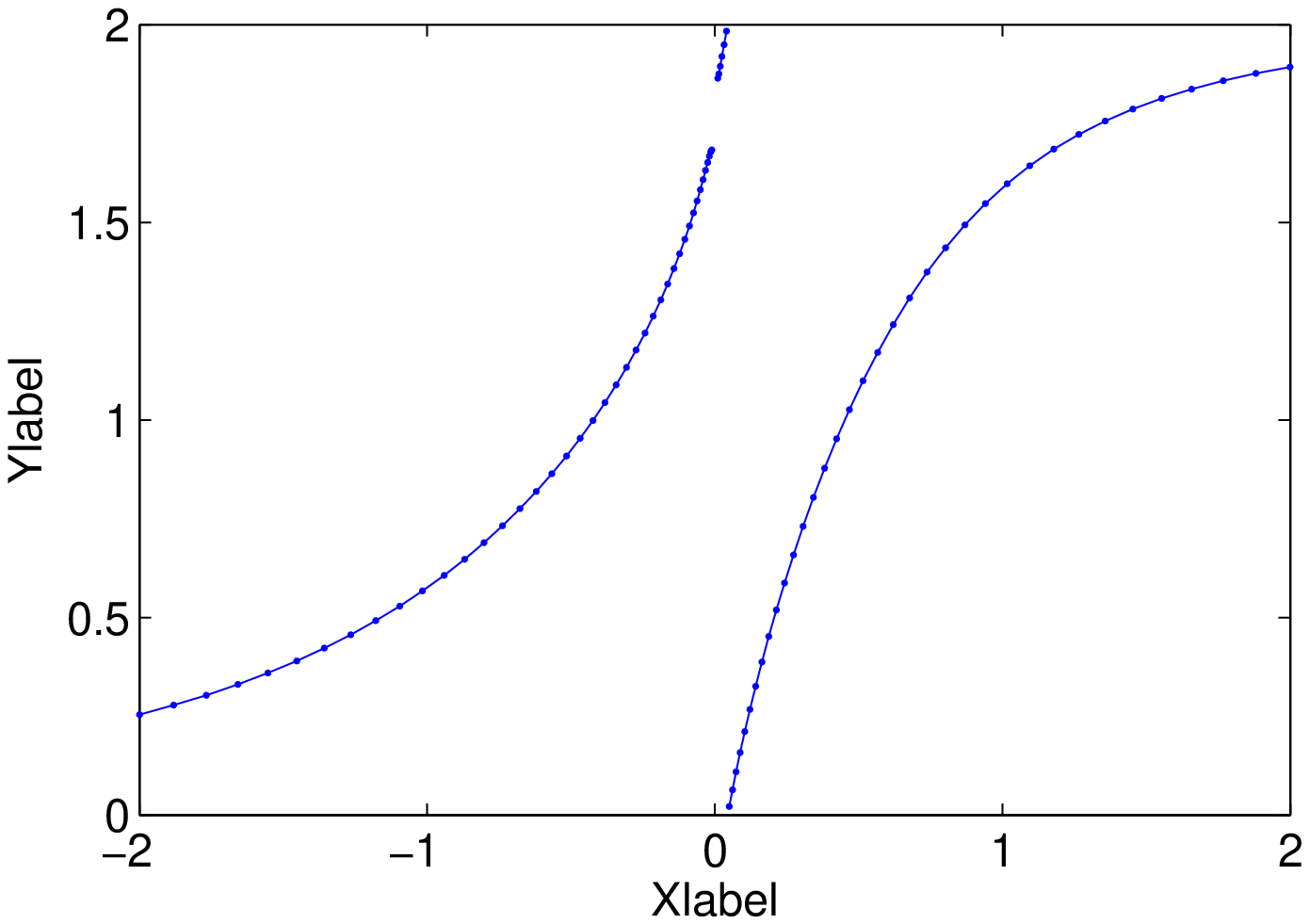}
  }\hspace{.5cm}%
  \subfigure{\label{subfig:7phase}
    \includegraphics[width=.1815\textwidth,height=.1727\textwidth]
      {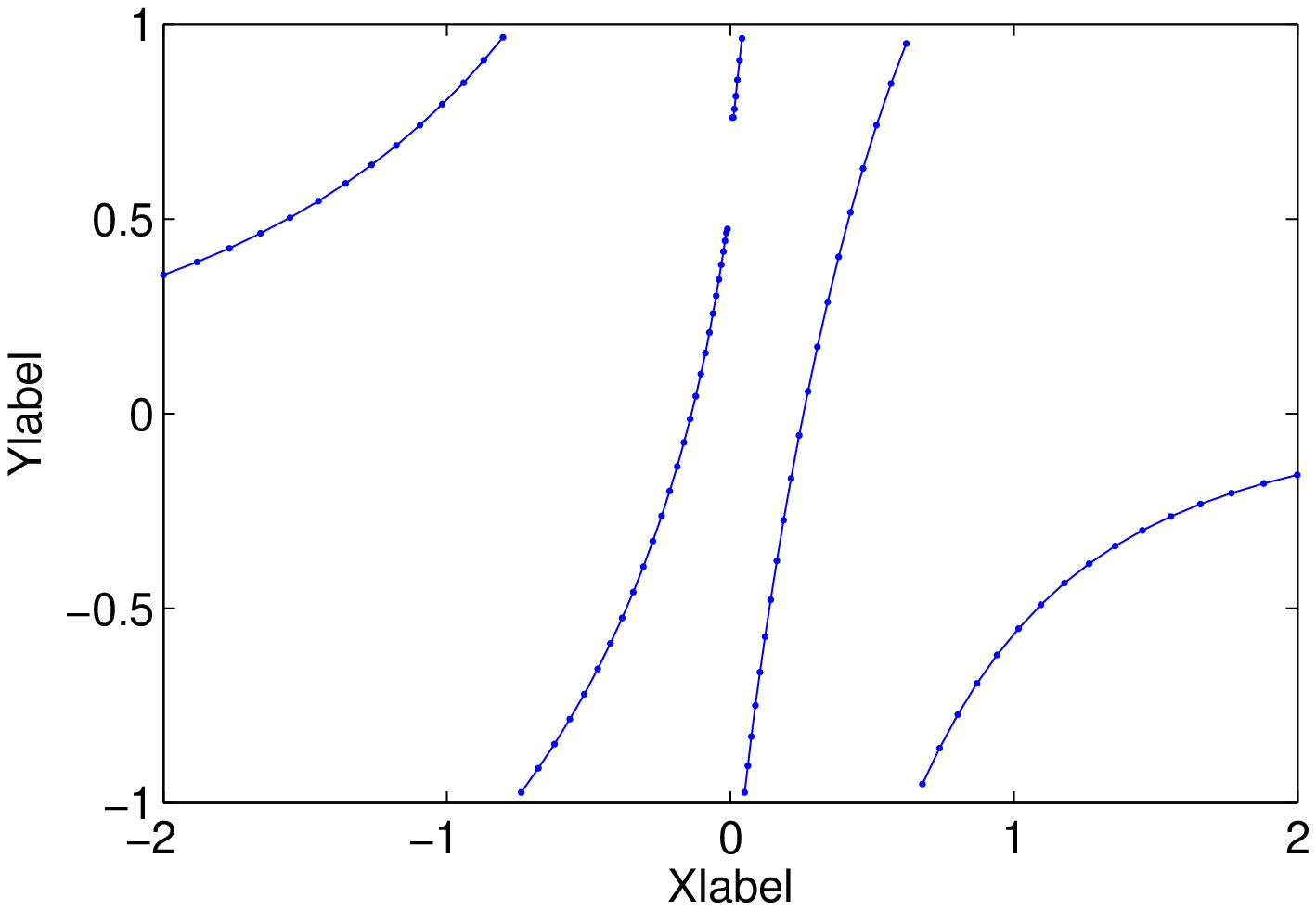}
  }\\[0.2cm]%
  \subfigure{\label{subfig:5joinedphase}
    \includegraphics[width=.2\textwidth,height=.18\textwidth]
      {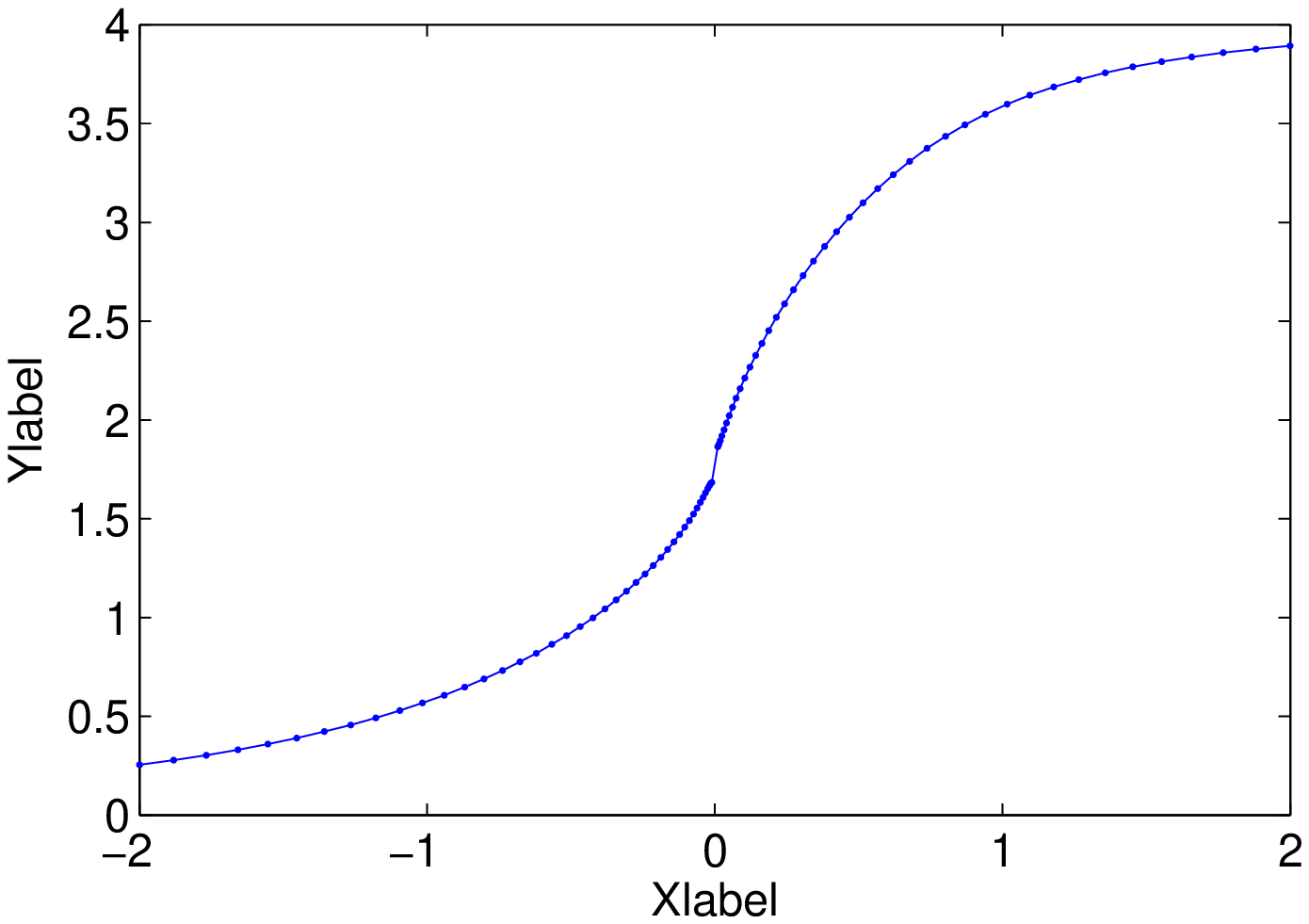}
  }\hspace{.25cm}%
  \subfigure{\label{subfig:7joinedphase}
    \includegraphics[width=.2\textwidth,height=.18\textwidth]
      {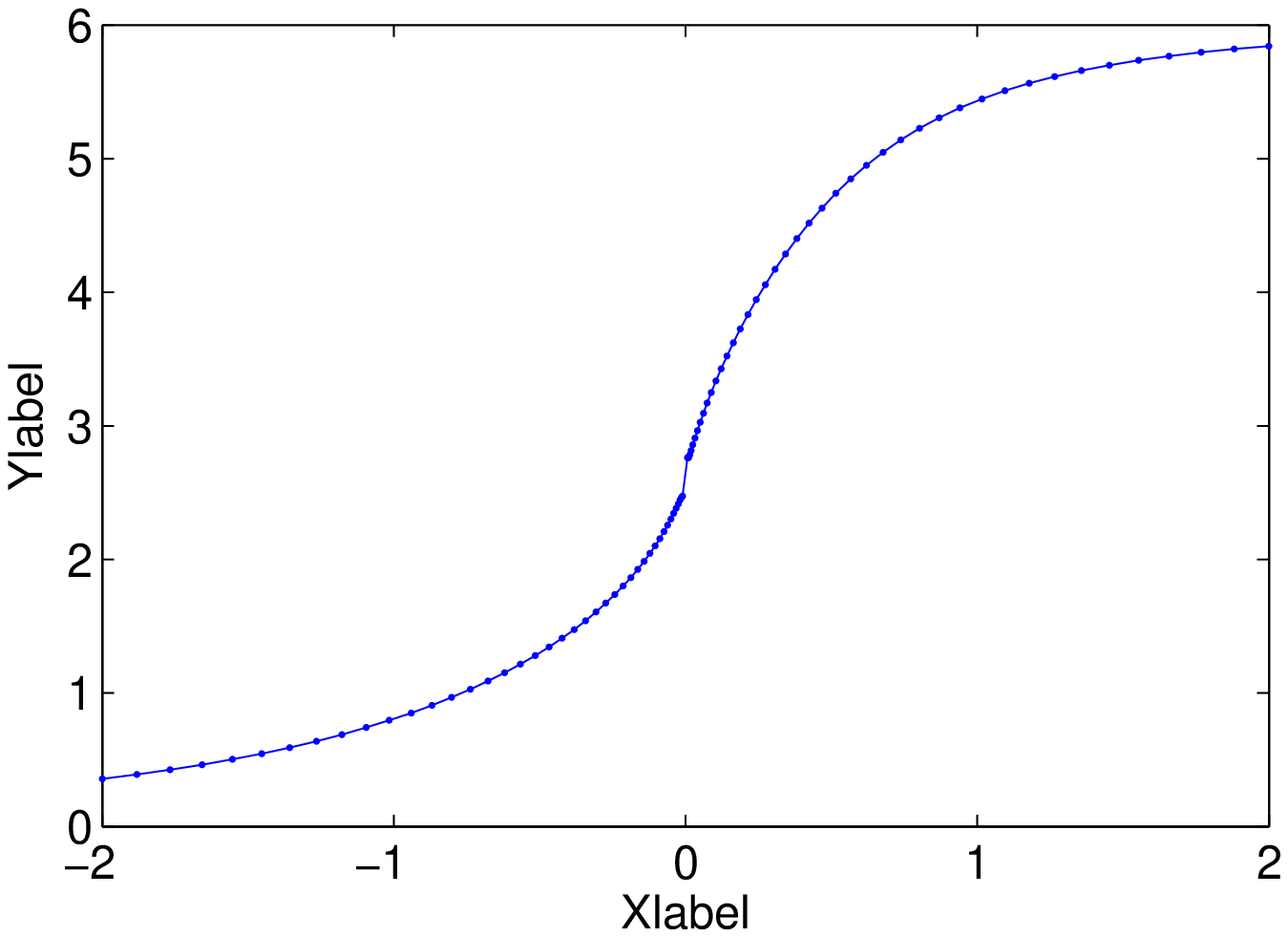}
  }\\[0.2cm]%
  \psfrag{Ylabel}[bc][bc]{$|\mathcal{C}|$}
  \subfigure{\label{subfig:5abscnum}
    \includegraphics[width=.18\textwidth,height=.17\textwidth]
      {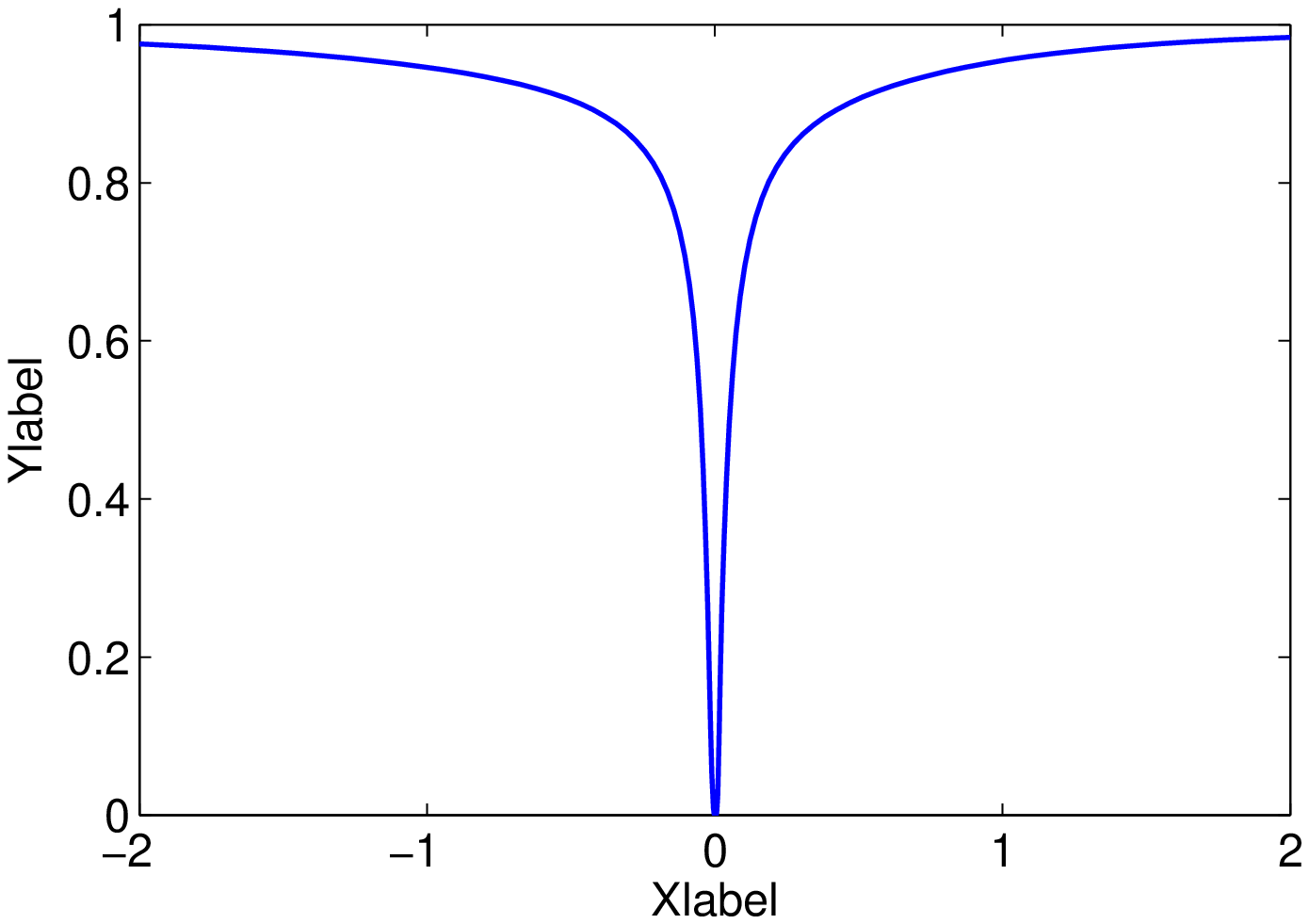}
  }\hspace{.7cm}%
  \subfigure{\label{subfig:7abscnum}
    \includegraphics[width=.18\textwidth,height=.17\textwidth]
      {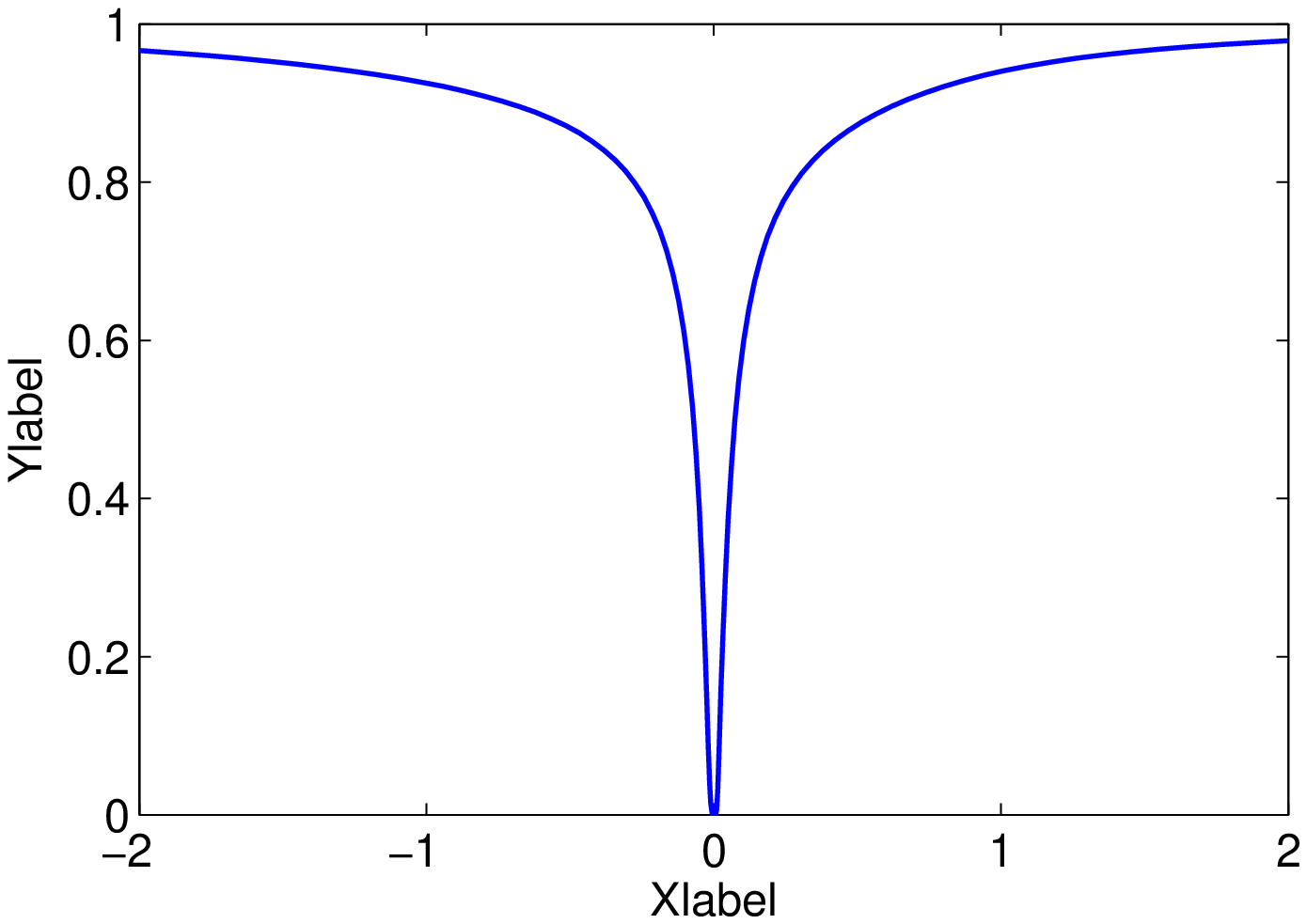}
  }\\[0.2cm]%
  \caption{
These plots depict the Bargmann invariants for spin-$1$ chains of
lengths $N=5$ (left-hand column) and $N=7$ (right-hand column),
and allow for direct comparison and contrast with the spin-$1/2$
case of Fig.~\ref{fig:spinonehalf}. For a detailed description of
the results the reader is kindly referred to the text in
section~\ref{section:spinone}.
  }
  \label{fig:spinone}
\end{figure}

The results presented in Fig.~\ref{fig:spinone} derive from
simulations that are entirely analogous to those that gave rise to
Fig.~\ref{fig:spinonehalf}, but they refer to the spin-$1$ chain
model, rather than the spin-$1/2$ model, Eq.~\ref{eq:generalham}.
The Bargmann invariants  are shown for spin-$1$ chains of lengths
$N=5$ (left-hand column) and $N=7$ (right-hand column), and allow
for direct comparison and contrast with the spin-$1/2$ case of
Fig.~\ref{fig:spinonehalf}. The circuit is composed of
$\mathcal{N} = 100$ vertices. Again, the uppermost row shows the
trajectory traced out in the complex plane by the Bargmann
invariant as the spin coupling strength is varied from $J_c - 2r$
to $J_c + 2r$. The second row from the top shows the corresponding
Bargmann phase; notice that points in the immediate vicinity of
the critical point have been disregarded as the phase of a complex
number becomes increasingly prone to error as the origin of the
complex plane is approached. The next row pictures the overall
`extent' of the phase, with the individual pieces of the graph
having been joined up end-to-end, as described in
section~\ref{section:Results}. Finally, the magnitude of the
Bargmann invariant is shown, and is clearly seen to vanish in the
vicinity of the critical point.

Qualitatively, the results for spin-$1$ chains are very similar to
those of spin-$1/2$ chains. In particular, the magnitude of the
Bargmann invariant still represents a signature for the critical
point. Of note is also the fact that the `joined-up' Bargmann
phase extends twice as high for spin-$1$ chains as it does for
spin-$1/2$ chains, for the same number of particles. Again, this
can be understood from Berry's solid angle result for
non-interacting particles, where the geometric phase is
proportional to the spin dimension.

\section{Summary and Discussion}
\label{section:Summary}
In summary, we have shown how the magnitude and `speed' of the
Bargmann invariant can act as a signature of the critical point for
the finite one-dimensional spin chain model considered. This stands
in stark contrast to the Bargmann phase, which is oblivious to the
phase transition as long as the spin chain length remains finite.

It is conjectured that the signatures presented here in the
context of a specific spin chain model will uphold their validity
also in a wider, more general setting. It would be worthwhile to
conduct experimental investigations with a view to testing this
conjecture; similarly, any theoretical proof of the proposed
conjecture would be of considerable interest.

Finally, it is worth stressing that the observed signatures of
criticality are very likely a direct consequence of the finite
`speed' with which the circuit is traversed in parameter space,
i.e. a result of the discrete nature of the circuit. No matter how
finely spaced are the neighbouring vertices on the circuit, the
adiabatic approximation will always break down in the immediate
vicinity of the critical point, causing Landau-Zener tunnelling
effects~\cite{LandauZener} to take centre stage. In the limit of a
smooth, continuous circuit, the observed signatures of criticality
would likely disappear altogether.

\begin{acknowledgments}
The authors would like to acknowledge discussions with K.
Audenaert at early stages of this project, as well as helpful
comments by D. Gross and K. Kieling. We thank L. Reuter for her
invaluable help and patience in creating Fig.~\ref{fig:Bloch2} of
this paper.

This work was funded in part by Hewlett-Packard Ltd. via an EPSRC
CASE award, by QIP IRC with support from the EPSRC (GR/S82176/0),
by the EU Integrated Project Qubit Applications (QAP), which is
funded by the IST directorate as contract no. 015848, by the
Alexander von Humboldt Foundation and the The Royal Society.
\end{acknowledgments}

%%%%%%%%%%%%%%%%%%%%%%%%%%%%%%%%%%%%%%%%%%%%%%%%%%%%%%%%%%%%%%
%%%%%%%%%%%%%%%%%%%%%%%%%%%%%%%%%%%%%%%%%%%%%%%%%%%%%%%%%%%%%%

%%%%%%%%%%%%%%%%%%%%%%%%%%%%%%%%%%%%%%%%%%%%%%%%%%%%%%%%%%%%%%%


\begin{thebibliography}{99}

\bibitem{Vojta} M.\ Vojta, Rep.\ Prog.\ Phys.\ {\bf 66}, 2069
(2003).

\bibitem{EntMeasures} M.B.\ Plenio and S.\ Virmani, Quant.\ Inf.\
Comp.\ {\bf 7}, 1 (2007).

\bibitem{OsterlohEtAl}
A.\ Osterloh , L.\ Amico, G.\ Falci and R.\ Fazio, Nature {\bf 416}
(2002) 608.

\bibitem{BlockEntIC}
%% Entanglement properties of the harmonic chain
K.\ Audenaert, J.\ Eisert, M.B.\ Plenio and R.F.\ Werner, Phys.\
Rev.\ A {\bf 66}, 042327 (2002);
%% Entropy, Entanglement, and Area: Analytical Results for Harmonic
%% Lattice Systems
M.B.\ Plenio, J.\ Eisert, J.\ Drei{\ss}ig and M.\ Cramer, Phys.\
Rev.\ Lett.\ {\bf 94}, 060503 (2005);
%% Entanglement-area law for general bosonic harmonic lattice systems
M.\ Cramer, J.\ Eisert, M.B.\ Plenio and J.\ Drei{\ss}ig, Phys.\
Rev.\ A {\bf 73}, 012309 (2006).

\bibitem{VidalLatorre}
G.\ Vidal, J.\ Latorre, E.\ Rico and A.\ Kitaev, Phys.\ Rev.\ Lett.\
{\bf 90}, 227902 (2003).

\bibitem{Keating}
%% Random Matrix Theory and Entanglement in Quantum Spin Chains
J.P.\ Keating and F.\ Mezzadri, Commun.\ Math.\ Phys.\ {\bf 252},
543 (2004).

%% quant-ph/0411080
%% Multipartite entanglement in quantum spin chains
\bibitem{Bruss}
D.\ Bruss, N.\ Datta, A.\ Ekert, L.C.\ Kwek and C.\ Macchiavello,
Phys.\ Rev.\ A {\bf 72}, 014301 (2005).

%% Multipartite entanglement characterization of a quantum phase transition
\bibitem{Costantini}
G.\ Costantini, P.\ Facchi, G.\ Florio and S.\ Pascazio,
quant-ph/0612098 (2006).

\bibitem{RelEnt}
V.\ Vedral, M.B.\ Plenio, M.A.\ Rippin and P.L.\ Knight, Phys.\
Rev.\ Lett.\ {\bf 78}, 2275 (1997); V.\ Vedral, M.B.\ Plenio, K.\
Jacobs and P.L.\ Knight, Phys.\ Rev.\ A {\bf 56}, 4452 (1997); V.\
Vedral and M.B.\ Plenio, Phys.\ Rev.\ A {\bf 57}, 1619 (1998)

\bibitem{Geometric}
T.-C.\ Wei and P.M.\ Goldbart, Phys.\ Rev.\ A {\bf 68}, 042307
(2003); T.-C. Wei, M. Ericsson, P.M. Goldbart and W.J. Munro, Quant.
Inf. Comp. {\bf 4}, 252 (2004).

%% Ground state overlap and quantum phase transitions
%% quant-ph/0512249.
\bibitem{Zanardi}
P.\ Zanardi and N.\ Paunkovi$\acute{\text{c}}$, Phys.\ Rev.\ E {\bf
74}, 031123 (2006).

\bibitem{Berry}
M.V.\ Berry, Proc.\ R.\ Soc.\ London, Ser.\ A {\bf 392}, 45 (1984).

\bibitem{Carollo1}
A.\ Carollo and J.K.\ Pachos, Phys.\ Rev.\ Lett. {\bf 95}, 157203
(2005).

\bibitem{Carollo2}
A.\ Carollo and J.K.\ Pachos, quant-ph/0602154.

\bibitem{Hamma}
A.\ Hamma, quant-ph/060209.

\bibitem{Resta}
R.\ Resta, J.\ Phys.: Condensed Matter {\bf 12} (2000).

\bibitem{Simon}
B.\ Simon, Phys.\ Rev.\ Lett. {\bf 51}, 2167-2170 (1983).

\bibitem{Bargmann}
V.\ Bargmann, J.\ Math.\ Phys.\ {\bf 5}, 862 (1964).

\bibitem{Pancharatnam}
S.\ Pancharatnam, Proc.\ Indian Acad.\ Sci.\ {\bf 44}, Sec.\ A, 247
(1956).

\bibitem{footnote1} As an aside, we remark that we only
became aware of the vanishing energy gap at $J = |B|/2$ in this
system as a direct result of observing a discontinuity in the
magnitude of the Bargmann invariant. Of course, this is very much
in the spirit of this investigation, where Bargmann invariants are
assumed to take on a diagnostic role for critical points, even if
in hindsight the existence of that critical point is easily
demonstrated analytically.

%% Quantum information and triangular optical lattices
\bibitem{Kay}
A.\ Kay , D.K.K.\ Lee, J.K.\ Pachos, M.B.\ Plenio, M.E.\ Reuter and
E.\ Rico, Opt.\ Spectrosc.\ {\bf 99}, 339 (2005).

\bibitem{footnote2} We note in passing that we have also developed
matrix product state algorithms in an attempt to boost the number
of particles we can simulate. However, these analyses have not
succeded in producing qualitatively new results, over and obove
those presented here.

\bibitem{LandauZener}
L.D.\ Landau, Phys.\ Z.\ Sowjetunion {\bf 2}, 46 (1932); C.\ Zener,
Proc.\ R.\ Soc.\ London, Ser.\ A {\bf 137}, 696 (1932).

\end{thebibliography}
\end{document}